\documentclass[preprint]{elsarticle}
\usepackage[width=6.5in]{geometry}

\usepackage{setspace, parskip, xcolor, breakcites, hyphenat, graphicx, epstopdf, amsmath, amssymb, amsfonts, array,
	verbatim, url, subcaption, booktabs, multirow, etoolbox,siunitx, etoolbox, tikz,stackengine, textcomp,
	gensymb, siunitx,tabularx, array}
\usepackage[normalem]{ulem}
\newcolumntype{H}{>{\setbox0=\hbox\bgroup}c<{\egroup}@{}}
\usepackage[toc,page,titletoc]{appendix}
\usepackage[inline]{enumitem}
\usepackage[version=4]{mhchem}
\usepackage[T1]{fontenc}
\usepackage[font=footnotesize,skip=2pt]{caption}
\usepackage[version=4]{mhchem}
\definecolor{srm}{HTML}{034da1}
\definecolor{cof}{RGB}{219,144,71}
\definecolor{pur}{RGB}{186,146,162}
\definecolor{greeo}{RGB}{91,173,69}
\definecolor{greet}{RGB}{52,111,72}
\definecolor{mplc0}{HTML}{1f77b4}
\definecolor{mplc1}{HTML}{ff7f0e}
\definecolor{mplc2}{HTML}{2e7d32}
\definecolor{mplc3}{HTML}{d32f2f}
\definecolor{mplc4}{HTML}{9467bd}
\definecolor{mplc5}{HTML}{8c564b}
\definecolor{mplc6}{HTML}{e377c2}
\definecolor{mplc7}{HTML}{7f7f7f}
\definecolor{mplc8}{HTML}{bcbd22}
\definecolor{mplc9}{HTML}{17becf}
\definecolor{Astral}{HTML}{1F77B4}
\definecolor{BG80}{HTML}{37474f}
\definecolor{Green}{HTML}{4CA540}
\definecolor{Red}{HTML}{F44336}
\usepackage[colorlinks=true, linkcolor=mplc0, urlcolor=mplc3, filecolor=mplc0, citecolor=mplc2,
  pdfstartview=FitV, pdftitle={}, pdfauthor={}, pdfsubject={}, pdfkeywords={}, pdfpagemode={},
  bookmarksopen=true, breaklinks]{hyperref}

\newif\ifhighlight     % round-1 markup (\rev, red)
\newif\ifhighlightB    % round-2 markup (\revB, violet)
\highlightfalse        % round 1 off -> prints as normal black text
\highlightBfalse        % round 2 on  -> prints violet

\newcommand{\rev}[1]{%
  \ifhighlight
    \textcolor{mplc3}{#1}
  \else
    #1
  \fi
}

\newcommand{\revB}[1]{%
  \ifhighlightB
    \textcolor{mplc3}{#1}
  \else
    #1
  \fi
}

\journal{jmmm}

\usepackage[capitalise,nameinlink]{cleveref}
\crefname{supp}{Supplement}{Supplements}

\newcommand{\mfr}{\ce{Fe_{1-x}Mn_{x}Rh}}
\newcommand{\cfr}{\ce{Fe_{1-x}Co_{x}Rh}}
\newcommand{\zfr}{\ce{Fe_{1-x}(Mn/Co)_{x}Rh}}
\date{\today}
\begin{document}

\begin{frontmatter}

	\title{$d$-band filling dictates magnetic stability in Mn- and Co-substituted FeRh alloys}

	\author{Greeshma R}
	\author{Rudra Banerjee\corref{cor1}}
	\ead{rudrab@srmist.edu.in}

	\cortext[cor1]{Corresponding author}

	\affiliation{organization={Department of Physics and Nanotechnology, SRM Institute of
				Science and Technology}, addressline={Kattankulathur}, city={Chennai}, postcode={603203},
		state={Tamil Nadu}, country={India}}

	\begin{abstract}
		The composition-dependent magnetic properties of B2-ordered \zfr~alloys with substitutional disorder on the Fe sublattice are
		investigated using first-principles calculations within the coherent potential approximation. By systematically substituting Mn
		and Co on the Fe sublattice, we establish $d$-band filling as the primary control parameter governing magnetic stability in this
		itinerant system. Mn substitution (hole doping) shifts the Fermi level into the minority-spin bonding states, driving a collapse
		of spin polarization (crossing zero at $x \approx 0.5$) and the emergence of competing antiferromagnetic interactions
		($\eta_\mathrm{Mn} < 0$). Even though the ferromagnetic configuration remains energetically well separated from the G-type
		AFM-II configuration across the studied range ($\Delta E$ up to $\sim$0.35~eV/atom), this exchange competition drives an
		``itinerant magnetic softness'' that suppresses the Curie temperature by $\sim$450~K---a finite-temperature instability set by
		the near-cancellation of competing exchange interactions rather than by AFM--FM energy proximity. In contrast, Co substitution
		(electron doping) acts as a ``magnetic hardener'' by pinning the Fermi level within the majority-spin pseudogap, preserving high
		spin polarization ($|P| \approx 0.75$) and stabilizing ferromagnetic exchange across the full composition range. These results
		show that tuning the Fermi level relative to the pseudogap provides a systematic, microscopic framework for controlling magnetic
		stability in B2-ordered itinerant magnets, distinct from simple magneto-volume models.
	\end{abstract}


\end{frontmatter}

% Main text
% \input{ferh}

\section{Introduction}
Magnetic order in itinerant transition-metal alloys is set by the filling of the $3d$ band: the position of the Fermi level $E_F$ within the spin-resolved density of states fixes the local moments, the sign and range of the interatomic exchange interactions $J_{ij}$, and ultimately the magnetic ordering temperature \cite{Dolgikh_2025,Singh2023,Rumiantsev_2024}. Because the valence electron count per atom ($e/a$) controls this filling, substitutional alloying of Fe with transition metals of differing valence is a direct lever on ferromagnetism---a principle codified in the classical Slater-Pauling curve \cite{Pauling1938,Slater1936} and still central to the compositional design of magnetic alloys \cite{Galanakis2002,Kubler2000}. Doping Fe with Mn (hole doping, reduced $Z_\text{val}$) or Co (electron doping, increased $Z_\text{val}$) shifts the band occupation in opposite directions, offering a clean handle on how $E_F$ redistributes $J_{ij}$ across coordination shells \cite{Liechtenstein1987,Skomski2008,Ullah_2024}.

The near-equiatomic FeRh binary is an ideal testbed for this filling-controlled magnetism \cite{Dolgikh_2025,Rumiantsev_2024,Greeshma_2022}. It crystallizes in the B2 (\ce{CsCl}-type) structure and undergoes a first-order antiferromagnetic (AFM)-to-ferromagnetic (FM) metamagnetic transition near $320$--$370$~K \cite{Kouvel1962,Kudrnovsky2015}, accompanied by a $\sim$1\% volume expansion and the emergence of an induced Rh moment ($\sim$1.0~$\mu_B$) through spin-dependent Fe--Rh $d$--$d$ hybridization \cite{Polesya2016}. This transition underpins proposals for magnetocaloric cooling \cite{Nikitin1990,Annaorazov1996} and antiferromagnetic spintronic memory \cite{Marti2014}, and its temperature is strongly tunable by composition. The broad $4d$ bands of Rh mediate the exchange between Fe moments while the Fe sublattice carries the dominant magnetic response, so that substitution on the Fe site directly reshapes the exchange network.

Quite generally, AFM order is favoured near half-filling of the $d$ band whereas FM order prevails toward the empty- or filled-band limits \cite{Teraoka1990}, so that shifting $E_F$ through compositional modification \cite{Kuo2002} provides a route to tune the ordering. Experimentally, both Mn and Co substitution strongly suppress the metamagnetic transition temperature \cite{Joshi2025,Polesya2016} and alter the magnetization \cite{Nikitin1990}. On the theoretical side, the coherent potential approximation (CPA) \cite{CPA_Soven} has long been used to treat substitutional disorder in magnetic alloys, including Heusler systems where atomic substitution redistributes $J_{ij}$ across coordination shells \cite{Huang_2022}, and first-principles studies confirm that band-filling effects govern magnetism through the Fermi-level position and the spin-dependent density of states \cite{Rahman_2010,van_Schilfgaarde_1999}. \revB{For FeRh itself, first-principles alloy studies have concentrated on dilute substitution and its effect on the metamagnetic transition---mapping how impurities shift the AFM--FM energy balance and the transition temperature~\cite{Kudrnovsky2015,Polesya2016}---rather than on the fate of the exchange network at high substitution levels.}%R1.1

What remains unresolved is how $d$-band filling quantitatively regulates the \emph{site-resolved} competition between ferromagnetic and antiferromagnetic exchange pathways in B2-ordered FeRh under high-concentration substitutional disorder, and in particular how the contrasting roles of Mn hole doping and Co electron doping play out across the full composition range. The bulk Slater-Pauling dependence of the moment on $e/a$ is well established, but it does not by itself reveal which exchange pathways stiffen or soften as the band is filled, nor why nominally similar $3d$ substituents should drive opposite magnetic fates.

Here we address this gap with first-principles CPA calculations of the electronic structure, magnetic moments, and exchange
interactions of \zfr~alloys across $x = 0.1$--$0.8$, combined with an explicit Liechtenstein exchange analysis
\cite{Liechtenstein1987}. Because the Liechtenstein formalism evaluates $J_{ij}$ from infinitesimal rotations about a magnetic
reference configuration, we take the FM phase---the state realized above the metamagnetic transition and the one relevant to the
applications above---as that reference, and ask how Mn and Co substitution reshape the exchange interactions within it. This
resolves the divergent dopant roles---Mn reducing and Co increasing the valence electron count---in the magnetic moments, Curie
temperature $T_C$, and exchange competition, and traces their electronic-structure origin to the motion of $E_F$ relative to the
majority-spin pseudogap. We find that targeted $d$-band manipulation tunes the magnetic ordering temperature over a $\sim$450~K range, providing a systematic, microscopic picture of compositional control of itinerant ferromagnetism in this system. \revB{The paired series thus provide a two-sided test of $d$-band-filling control---hole and electron doping of the same sublattice, treated within a single formalism; the site- and shell-resolved exchange sums identify which pathways stiffen or soften under each; and the ground-state energy difference $\Delta E$ is held separate from the finite-temperature exchange 	competition, two measures of stability that need not coincide.}%R1.1
% \revB{This combination---full-range treatment of both doping directions, stability analysis at the level of site- and shell-resolved exchange sums rather than the bulk moment alone, and explicit separation of the ground-state $\Delta E$ from the finite-temperature exchange competition---distinguishes the present work from earlier studies of substituted FeRh.}%R1.1

\section{Computational Methods}
The electronic structure and magnetic properties of \mfr~and \cfr~alloys were investigated using density functional theory (DFT) within the multiple-scattering Green's function formalism, treating the substitutional disorder on the Fe sublattice within the CPA as implemented in the Spin-Polarized Relativistic Korringa-Kohn-Rostoker (SPR-KKR) code~\cite{sprkkr,Ebert_2011}. The parent \ce{FeRh} compound crystallizes in the B2 (\ce{CsCl}-type) structure (space group $Pm\bar{3}m$), consisting of two interpenetrating simple cubic sublattices; in the disordered alloys, Mn or Co atoms randomly replace Fe at the $1a$ Wyckoff site with probability $x$, while the Rh sublattice ($1b$) remains fully ordered. Exchange-correlation effects were treated within the Perdew-Burke-Ernzerhof (PBE) generalized gradient approximation~\cite{Perdew_1996}. \revB{No Hubbard-$U$ correction was applied: FeRh is an itinerant $3d$--$4d$ intermetallic with band-like, strongly hybridized $3d$ states, and its quantitative first-principles description---ground-state energetics, metamagnetic transition, exchange interactions, and substitutional magnetism---has been established with semi-local functionals in agreement with experiment~\cite{Staunton2014,Kudrnovsky2015,Polesya2016,Tran2022}. A recent systematic benchmark of exchange-correlation functionals for FeRh~\cite{Pandey_2025} finds the PBE-derived exchange interactions to closely reproduce the experimental ordering temperature while beyond-GGA treatments strongly overestimate it; no ab initio $U$ has been established for this metallic system, and an ad hoc $U$ would introduce an uncontrolled parameter into the Fermi-level position and the exchange interactions on which the present analysis rests.} Brillouin-zone integrations used a $k$-mesh of 3000 points in the irreducible wedge, and the self-consistent cycle was converged to an energy tolerance of $10^{-5}$~Ry. Calculations were performed in the spin-polarized, scalar-relativistic framework within the muffin-tin geometry with an angular-momentum cutoff $l_{\mathrm{max}} = 2$ ($s$, $p$, $d$ channels), the established setting for $3d$--$4d$ transition-metal alloys in the KKR-CPA framework~\cite{Kudrnovsky2015,Turek_2006}.

Equilibrium lattice parameters were obtained from total-energy calculations for lattice scalings between 94\% and 106\% of an initial value taken from our previous work~\cite{Greeshma_2022}; the resulting energy-volume data were fitted to a quadratic polynomial whose minimum identifies the optimized lattice constant, with the optimized ordered-FeRh value serving as the initial guess for the non-stoichiometric compositions. The optimized lattice parameters for all compositions are reported in \cref{tab:struct-mn,tab:struct-co}.

\revB{Within the single-site CPA the substitutional disorder is treated on the ideal B2 lattice, so static local displacements around the dopant atoms are neglected; because such displacements average to zero over impurity configurations, the ensemble-averaged positions remain the ideal sites, and what is omitted is the static fluctuation of bond lengths about that average. This approximation is expected to be mild here: Mn, Fe, and Co are closely size-matched $3d$ neighbours---the optimized lattice parameter varies by less than 1\% across the Mn series and contracts smoothly, in a Vegard-like manner, by $\approx$1.2\% across the Co series (\cref{tab:struct-mn,tab:struct-co})---and experimentally the substituted alloys retain the B2 structure, the Mn series in fact interpolating between the two isostructural B2 compounds \ce{FeRh} and \ce{MnRh}~\cite{Horky2022,Joshi2025,Seo_2024}. The dominant structural response---the volume---is included through the per-composition lattice optimization, and the magneto-volume analysis of \cref{sec:vol-chem} shows the magnetic trends to follow the valence electron count rather than the lattice parameter, indicating that residual local relaxations, a higher-order structural effect, do not alter the band-filling mechanism established here~\cite{Ebert_2011,Turek_2006}.}%R1.3

The spin polarization $P$ at the Fermi level was evaluated as
\begin{equation}
	P = \frac{N_{\uparrow}(E_{\mathrm{F}}) - N_{\downarrow}(E_{\mathrm{F}})}{N_{\uparrow}(E_{\mathrm{F}}) + N_{\downarrow}(E_{\mathrm{F}})},
	\label{eq:spinpol}
\end{equation}
where $N_{\uparrow}(E_{\mathrm{F}})$ and $N_{\downarrow}(E_{\mathrm{F}})$ are the spin-resolved densities of states at $E_{\mathrm{F}}$ for the majority and minority channels, respectively.

To analyze the magnetic stability, the itinerant electronic structure is mapped onto an effective classical Heisenberg Hamiltonian,
\begin{equation}
	H = - \sum_{\langle i,j \rangle} J_{ij} \, \mathbf{S}_i \cdot \mathbf{S}_j,
	\label{eq:heisenberg}
\end{equation}
with the convention that $J_{ij} > 0$ favours ferromagnetic and $J_{ij} < 0$ antiferromagnetic alignment of the moments at sites $i$ and $j$. The exchange parameters were computed from the Liechtenstein (LKAG) formula~\cite{Liechtenstein_1984,Terasawa_2019},
\begin{equation}
	J_{ij} = \frac{1}{4\pi} \int \mathrm{d}\varepsilon\, f\!\left[\beta(\varepsilon - \varepsilon_{\mathrm{F}})\right] \, \mathrm{Im}\,\mathrm{Tr}\!\left( \hat{\Delta}_i \hat{T}^{\,ij}_{\uparrow} \hat{\Delta}_j \hat{T}^{\,ji}_{\downarrow} \right),
	\label{eq:jij}
\end{equation}
where $\hat{T}^{\,ij}_{\sigma}$ is the scattering-path operator between sites $i$ and $j$, $\hat{\Delta}_i = \hat{t}_{i\uparrow}
	- \hat{t}_{i\downarrow}$ is the exchange splitting of the single-site scattering matrix, $f$ is the Fermi-Dirac function with
$\beta = 1/k_{\mathrm{B}}T$, and $\varepsilon_{\mathrm{F}}$ is the Fermi level. The formula evaluates the energy cost of
infinitesimal spin rotations about a magnetic reference configuration; following the magnetic force theorem we adopt the FM state
as that reference throughout, consistent with the phase realized above the metamagnetic transition. \rev{The $J_{ij}$ extracted
	in this way are effective exchange parameters describing transverse spin rotations about the FM reference; they characterize the
	exchange topology of the FM state and the composition trends in $T_C$, not the energetics of the first-order metamagnetic AFM--FM
	transition itself.} %R2.1, R3.1, R3.3

The Curie temperature was estimated within the mean-field approximation (MFA)~\cite{Liechtenstein_1987,Witte_2016,Sokolovskiy_2012},
\begin{equation}
	k_{\mathrm{B}} T_{\mathrm{C}} = \frac{2}{3} \lambda_{\mathrm{max}},
	\label{eq:tc}
\end{equation}
where $\lambda_{\mathrm{max}}$ is the largest eigenvalue of the real-space exchange matrix. The MFA is known to overestimate absolute $T_C$ in itinerant systems by 20--30\% through its neglect of spin fluctuations~\cite{Zhuravlev_2015,Mohn1999,Kudrnovsky2015}, but it reliably reproduces concentration-dependent trends across a composition series~\cite{Turek_2006,Karunakaran_2025}; we therefore treat the computed $T_C$ values as upper bounds and base our conclusions on the Mn-suppression and Co-stabilization trends rather than on the absolute numbers.

Because the dopant- and host-centred exchange environments differ in disordered alloys, the exchange competition parameter $\eta$ was evaluated separately for $J_{ij}$ with (i) Mn or Co and (ii) Fe at the central site,
\begin{equation}
	\eta = \frac{\sum_{i,j} J_{ij}}{\sum_{i,j} \lvert J_{ij} \rvert},
	\label{eq:frustration}
\end{equation}
so that $\eta = +1$ denotes purely ferromagnetic coupling, $\eta = -1$ purely antiferromagnetic coupling, and $\eta \to 0$ strong competition between FM and AFM pathways. In the classical limit $\eta$ coincides with the conventional frustration parameter; we use ``competition'' rather than ``frustration'' to emphasize that here the cancellation is generated electronically, by the redistribution of $J_{ij}$ as the $d$ band is filled, rather than by geometric constraints on a fixed spin lattice.

\section{Results}
\label{sec:results}
The magnetic behaviour of Mn- and Co-substituted FeRh arises from the competition between itinerant ferromagnetic and antiferromagnetic exchange pathways~\cite{Moriya1985,Liechtenstein1987,Ryan_2013}. To verify that the ferromagnetic (FM) configuration is the appropriate reference for the exchange analysis, we compared its energy with that of the G-type (AFM-II) configuration, in which the transition-metal moments alternate along $[111]$---the ordering realized in the low-temperature phase of equiatomic FeRh.

\begin{figure*}[htbp]
	\centering
	\begin{subfigure}[b]{.34\textwidth}
		\includegraphics[width=0.9\textwidth]{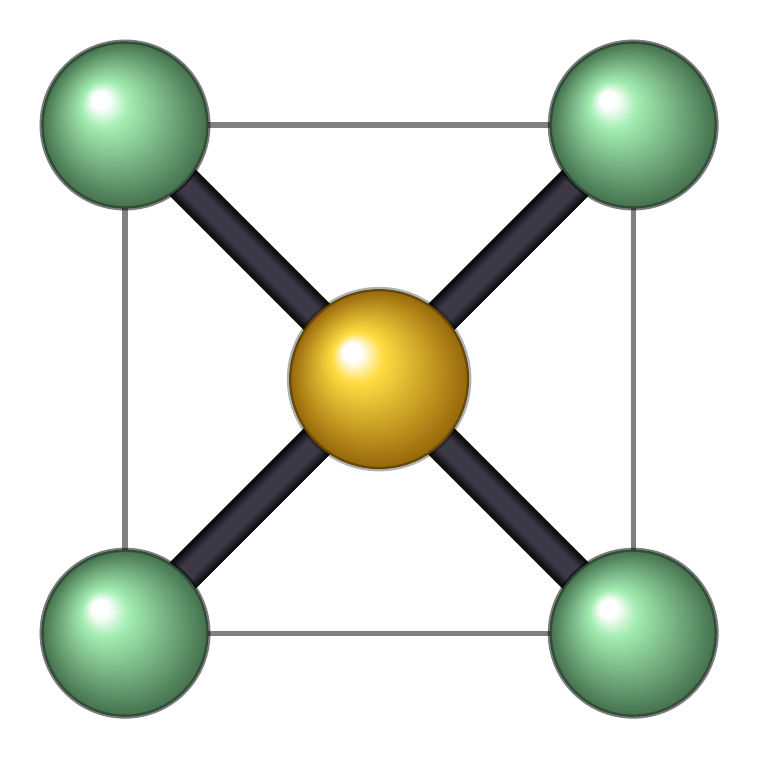}
		\caption{}
		\label{fig:structure}
	\end{subfigure}\hfill
	\begin{subfigure}[b]{.46\textwidth}
		\includegraphics[width=0.9\textwidth]{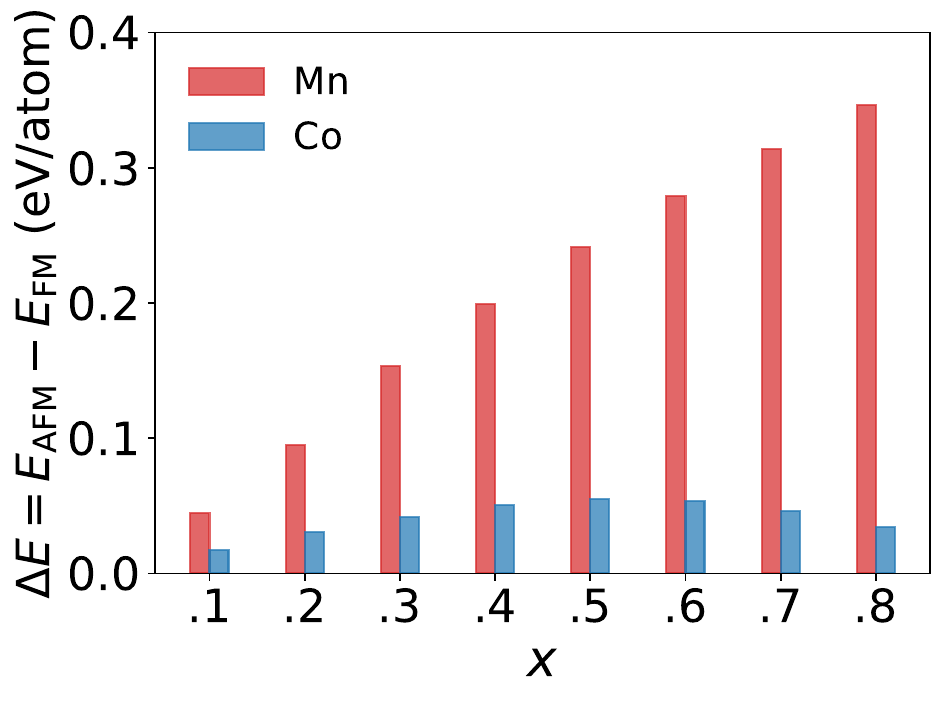}
		\caption{}
		\label{fig:deltaE}
	\end{subfigure}
	\caption{Structure and FM--AFM energetics of \zfr. (\subref{fig:structure})~B2 (\ce{CsCl}-type) crystal structure of \ce{FeRh}. Fe and Rh occupy two interpenetrating simple cubic sublattices ($1a$ and $1b$ Wyckoff sites, respectively); in the disordered alloys, Mn or Co randomly replaces Fe at the $1a$ site with probability $x$. (\subref{fig:deltaE})~AFM--FM energy difference $\Delta E = E_\mathrm{AFM} - E_\mathrm{FM}$ (eV/atom) for Mn- and Co-substituted FeRh. Positive values indicate that the FM configuration is lower in energy; it is the lower-energy state across the studied range $x = 0.1$--$0.8$. The corresponding numerical values are listed in \cref{tab:struct-mn,tab:struct-co}.}
\end{figure*}

The AFM--FM energy differences $\Delta E = E_\mathrm{AFM} - E_\mathrm{FM}$ are shown in \cref{fig:deltaE} and tabulated in \cref{tab:struct-mn,tab:struct-co}. Across the studied range $x = 0.1$--$0.8$, $\Delta E > 0$ for every composition, so the FM configuration is lower in energy than the G-type AFM-II configuration and constitutes a stable reference state for the Liechtenstein exchange analysis. In the Mn series $\Delta E$ grows monotonically, from $\sim$0.05~eV/atom at $x = 0.1$ to $\sim$0.35~eV/atom at $x = 0.8$, while in the Co series it remains comparatively modest ($\sim$0.02--0.06~eV/atom). We emphasize that this energetic ordering establishes the validity of the FM reference over the studied composition range and is not used to infer the absolute ground state of the parent $x = 0$ compound, whose metamagnetism is well documented~\cite{Kouvel1962,Kudrnovsky2015}; the central results below concern the exchange topology, which is independent of the AFM--FM energetics. Indeed, despite the large $\Delta E$ in the Mn series, those alloys exhibit a pronounced $T_C$ suppression (\cref{sec:jij}), demonstrating that the finite-temperature instability is governed by the exchange topology ($J_{ij}$, $\eta$) rather than by AFM--FM energy proximity.

The FM configuration therefore serves as the reference for extracting $J_{ij}$ via the Liechtenstein formalism and for the finite-temperature estimates. Within this reference all moments---including Mn---are aligned parallel; the consequences of the negative Mn-centred competition parameter $\eta_\mathrm{Mn}$ for the preferred Mn alignment are taken up in \cref{sec:discussion}.

\subsection{Electronic structure}
\label{sec:dos}
The evolution of the magnetic state in \zfr~is driven by the systematic $d$-band filling that moves the Fermi level $E_F$ through the spin-resolved density of states (DOS). In the parent B2 \ce{FeRh}, whose electronic structure we characterized previously~\cite{Greeshma_2022}, the majority ($\uparrow$) $d$-band is essentially filled and lies below $E_F$, so the majority-spin DOS at $E_F$ is low [$N_\uparrow(E_F) \approx 0.30$~states/eV; $x = 0$ row of \cref{tab:struct-mn}], whereas the minority ($\downarrow$) $d$-band straddles $E_F$ and remains high [$N_\downarrow(E_F) \approx 2.33$~states/eV]. The Fermi level is thus pinned in a majority-spin pseudogap, and the resulting strong spin-channel asymmetry sustains a high spin polarisation of minority character ($|P| \approx 0.77$, i.e.\ $P < 0$). Chemical substitution acts as precise $d$-band manipulation: Mn provides hole doping (reduced $Z_\text{val}$) and Co electron doping (increased $Z_\text{val}$), following generalized Slater-Pauling behaviour~\cite{Skomski2008,Mohn1999}.

% Integrated from SI: structural + electronic-structure data (former Tables S1, S2)
\begin{table*}[htbp]
	\centering
	\begin{tabular}{cccccc}
		\toprule
		\multirow{2}{*}{$x$} & \multirow{2}{*}{$a$ (\AA)}                   & \multirow{2}{*}{$E_F$ (eV)}
		                     & \multicolumn{2}{c}{DOS at $E_F$ (states/eV)} & \multirow{2}{*}{$\Delta E$ (eV/atom)}                                                   \\
		\cmidrule(lr){4-5}
		                     &                                              &                                       & $N_\uparrow(E_F)$ & $N_\downarrow(E_F)$ &       \\
		\midrule
		0.0                  & 3.0493                                       & 9.8188                                & 0.3009            & $-2.3274$           & ---   \\
		0.1                  & 3.0706                                       & 9.5459                                & 0.3287            & $-2.0817$           & 0.045 \\
		0.2                  & 3.0605                                       & 9.6859                                & 0.3669            & $-1.7948$           & 0.095 \\
		0.3                  & 3.0643                                       & 9.6256                                & 0.4265            & $-1.4719$           & 0.153 \\
		0.4                  & 3.0668                                       & 9.5577                                & 0.5414            & $-1.1003$           & 0.199 \\
		0.5                  & 3.0718                                       & 9.4428                                & 0.7476            & $-0.8110$           & 0.242 \\
		0.6                  & 3.0718                                       & 9.4278                                & 0.9418            & $-0.6707$           & 0.279 \\
		0.7                  & 3.0718                                       & 9.4271                                & 1.1018            & $-0.5767$           & 0.314 \\
		0.8                  & 3.0737                                       & 9.4123                                & 1.2317            & $-0.4994$           & 0.346 \\
		\bottomrule
	\end{tabular}
	\caption{Optimized lattice parameter $a$, Fermi energy $E_F$, spin-resolved DOS at $E_F$, and FM--AFM energy difference $\Delta E$ (eV/atom; plotted in \cref{fig:deltaE}) for the FM configuration of \mfr~across $x = 0$--$0.8$. The minority-channel DOS is reported with the conventional negative sign. Positive $\Delta E$ indicates the FM state is lower in energy.}
	\label{tab:struct-mn}
\end{table*}

\begin{table*}[htbp]
	\centering
	\begin{tabular}{cccccc}
		\toprule
		\multirow{2}{*}{$x$} & \multirow{2}{*}{$a$ (\AA)}                   & \multirow{2}{*}{$E_F$ (eV)}
		                     & \multicolumn{2}{c}{DOS at $E_F$ (states/eV)} & \multirow{2}{*}{$\Delta E$ (eV/atom)}                                                   \\
		\cmidrule(lr){4-5}
		                     &                                              &                                       & $N_\uparrow(E_F)$ & $N_\downarrow(E_F)$ &       \\
		\midrule
		0.0                  & 3.0493                                       & 9.8188                                & 0.3009            & $-2.3274$           & ---   \\
		0.1                  & 3.0424                                       & 9.8168                                & 0.3166            & $-2.3079$           & 0.017 \\
		0.2                  & 3.0418                                       & 9.7403                                & 0.3308            & $-2.3797$           & 0.031 \\
		0.3                  & 3.0342                                       & 9.7583                                & 0.3458            & $-2.4306$           & 0.042 \\
		0.4                  & 3.0305                                       & 9.7300                                & 0.3602            & $-2.5089$           & 0.051 \\
		0.5                  & 3.0267                                       & 9.7071                                & 0.3759            & $-2.5929$           & 0.055 \\
		0.6                  & 3.0230                                       & 9.6890                                & 0.3908            & $-2.6848$           & 0.054 \\
		0.7                  & 3.0192                                       & 9.6751                                & 0.4055            & $-2.7749$           & 0.046 \\
		0.8                  & 3.0142                                       & 9.6807                                & 0.4189            & $-2.8568$           & 0.034 \\
		\bottomrule
	\end{tabular}
	\caption{As \cref{tab:struct-mn}, for the FM configuration of \cfr~across $x = 0$--$0.8$. The $x = 0$ row is the shared pure-\ce{FeRh} parent (identical to \cref{tab:struct-mn}).}
	\label{tab:struct-co}
\end{table*}

\begin{figure*}[htbp]
	\centering
	% Row 1: Mn
	\begin{subfigure}[b]{0.32\textwidth}
		\includegraphics[width=\textwidth]{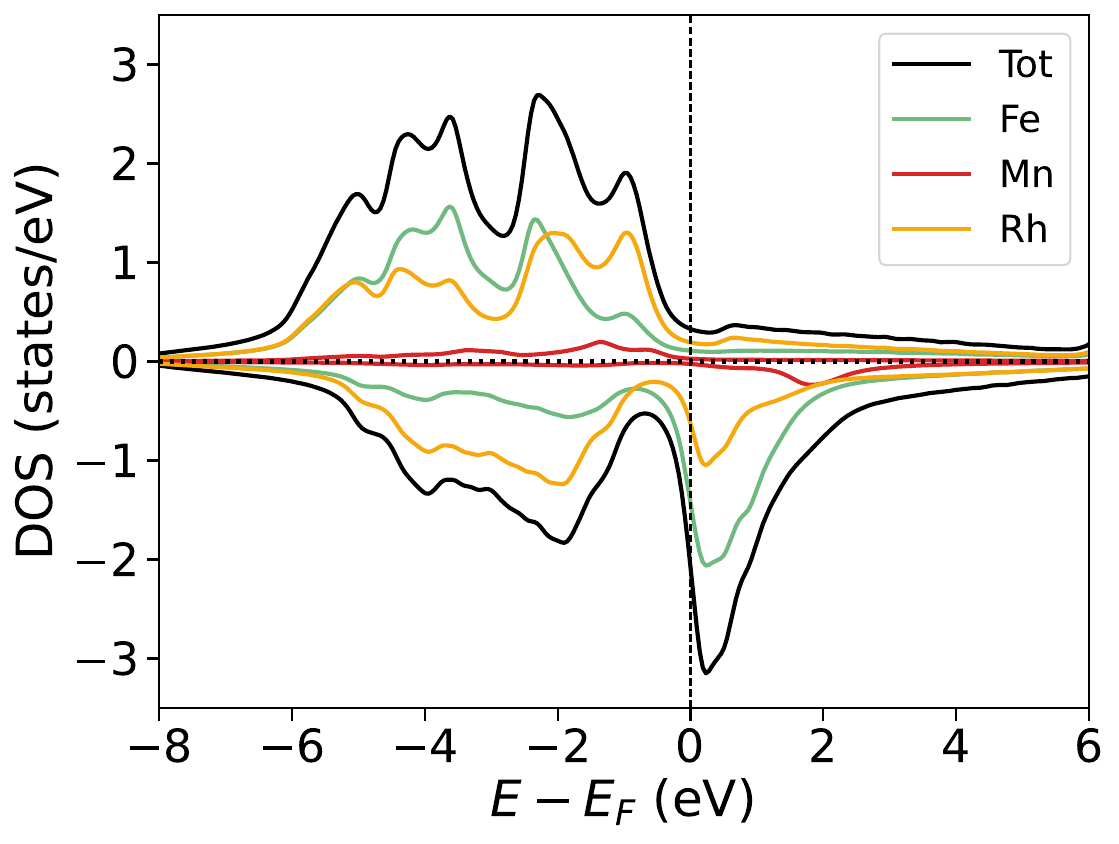}
		\caption{\ce{Fe_{0.9}Mn_{0.1}Rh}}
		\label{fig:dos_mn10}
	\end{subfigure}\hfill
	\begin{subfigure}[b]{0.32\textwidth}
		\includegraphics[width=\textwidth]{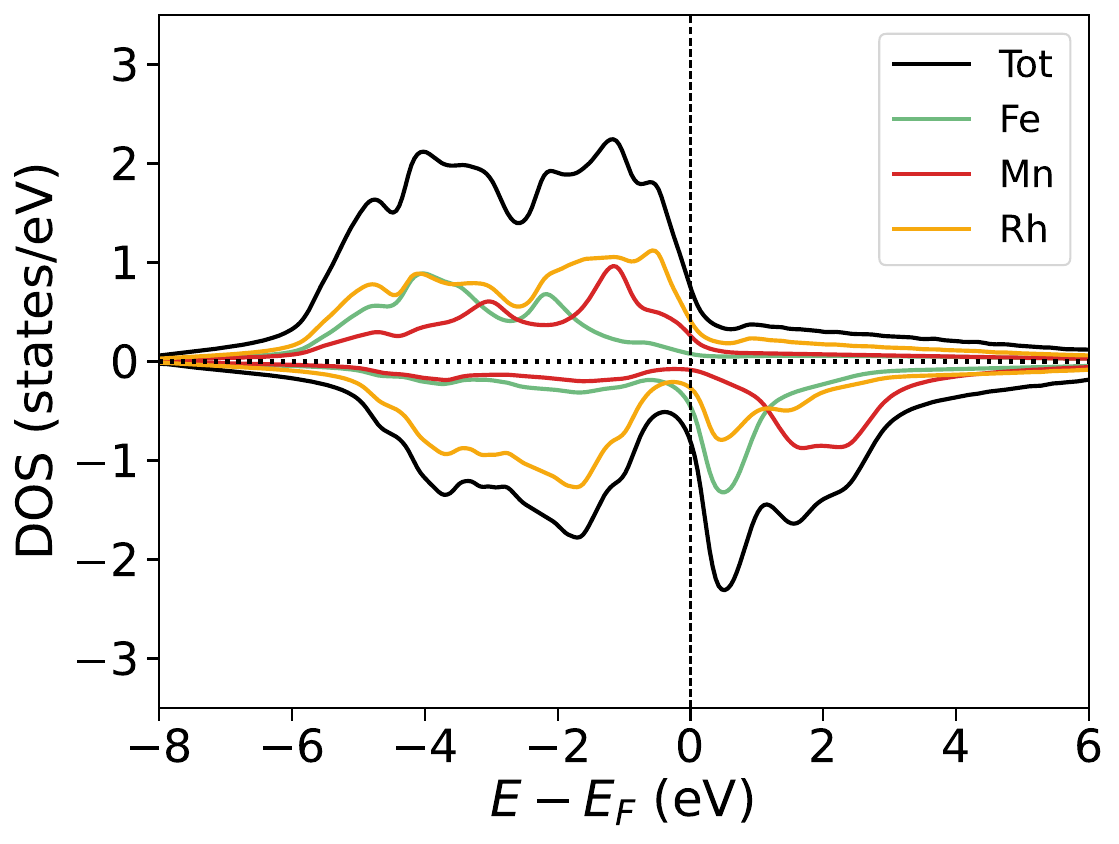}
		\caption{\ce{Fe_{0.5}Mn_{0.5}Rh}}
		\label{fig:dos_mn50}
	\end{subfigure}\hfill
	\begin{subfigure}[b]{0.32\textwidth}
		\includegraphics[width=\textwidth]{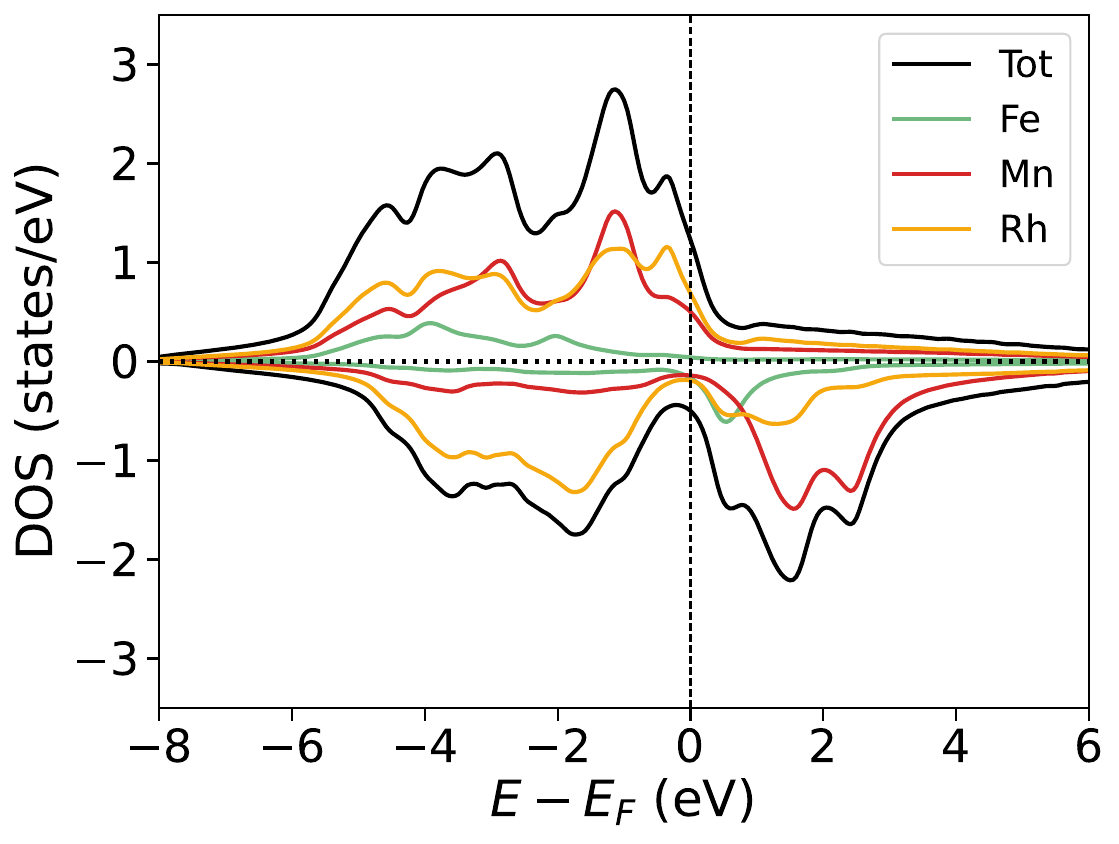}
		\caption{\ce{Fe_{0.2}Mn_{0.8}Rh}}
		\label{fig:dos_mn80}
	\end{subfigure}
	% Row 2: Co
	\begin{subfigure}[b]{0.32\textwidth}
		\includegraphics[width=\textwidth]{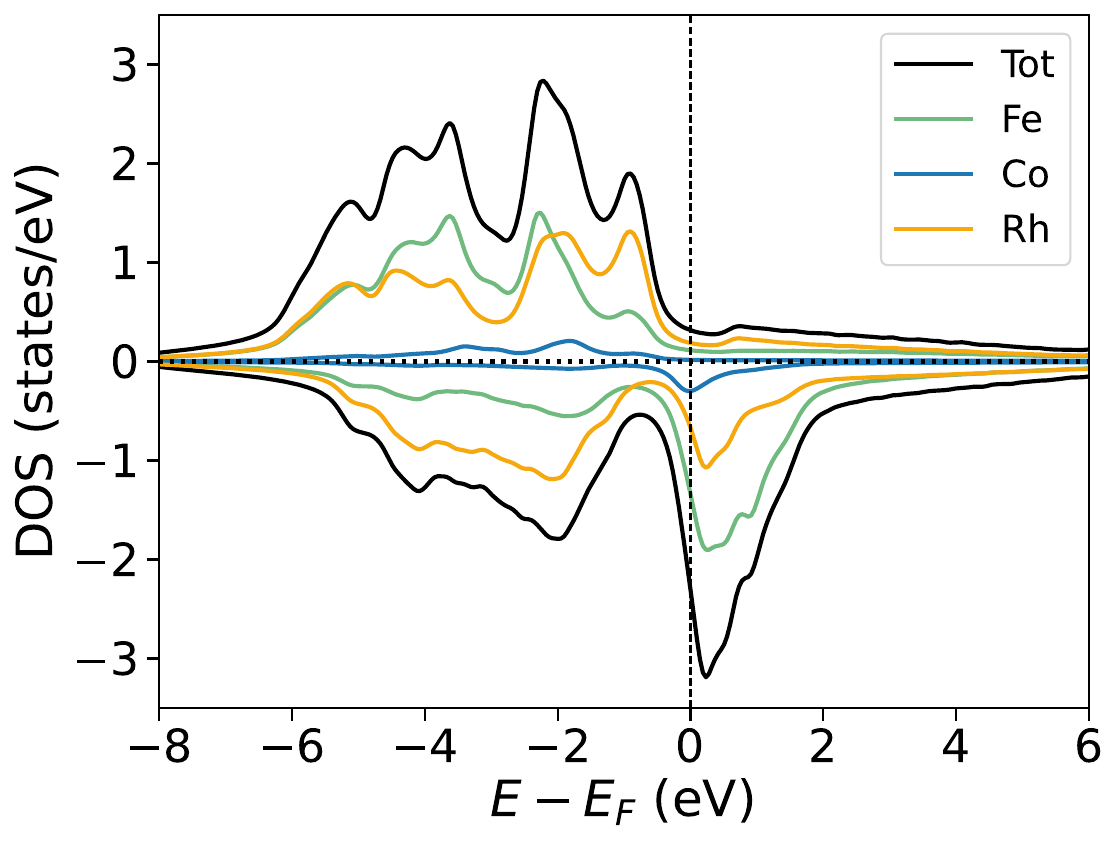}
		\caption{\ce{Fe_{0.9}Co_{0.1}Rh}}
		\label{fig:dos_co10}
	\end{subfigure}\hfill
	\begin{subfigure}[b]{0.32\textwidth}
		\includegraphics[width=\textwidth]{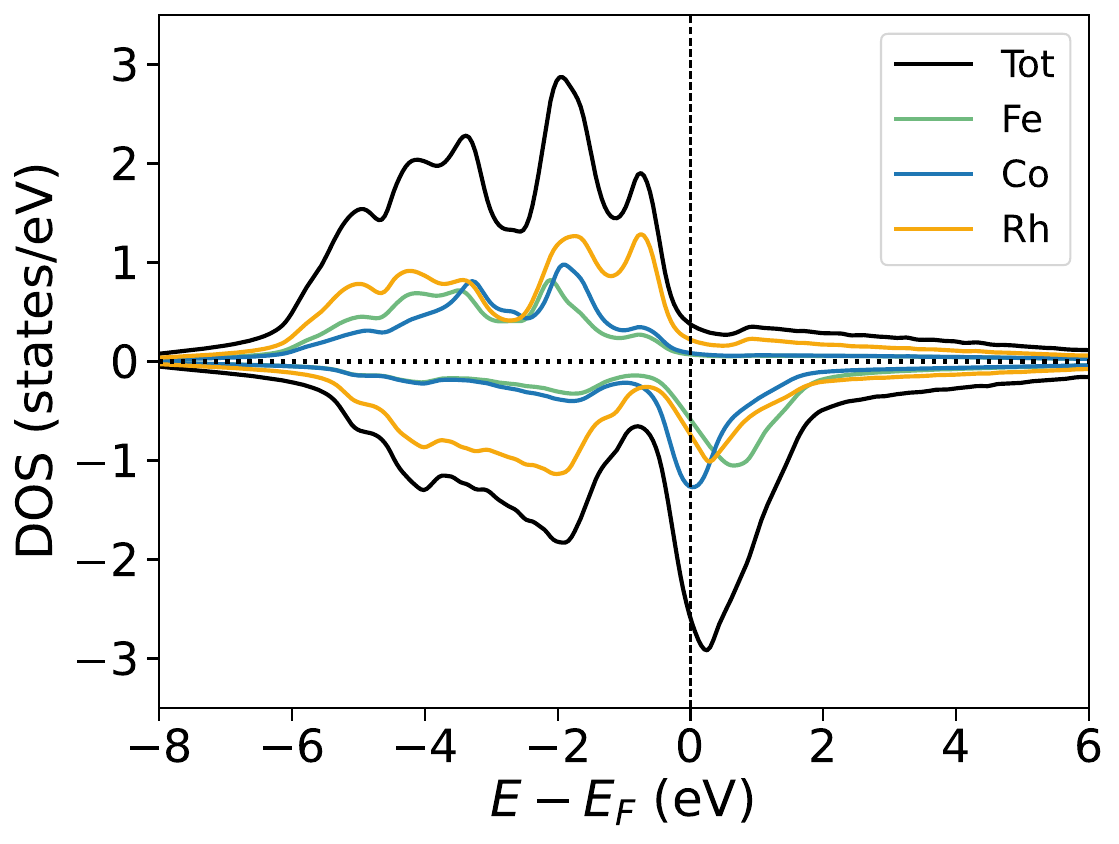}
		\caption{\ce{Fe_{0.5}Co_{0.5}Rh}}
		\label{fig:dos_co50}
	\end{subfigure}\hfill
	\begin{subfigure}[b]{0.32\textwidth}
		\includegraphics[width=\textwidth]{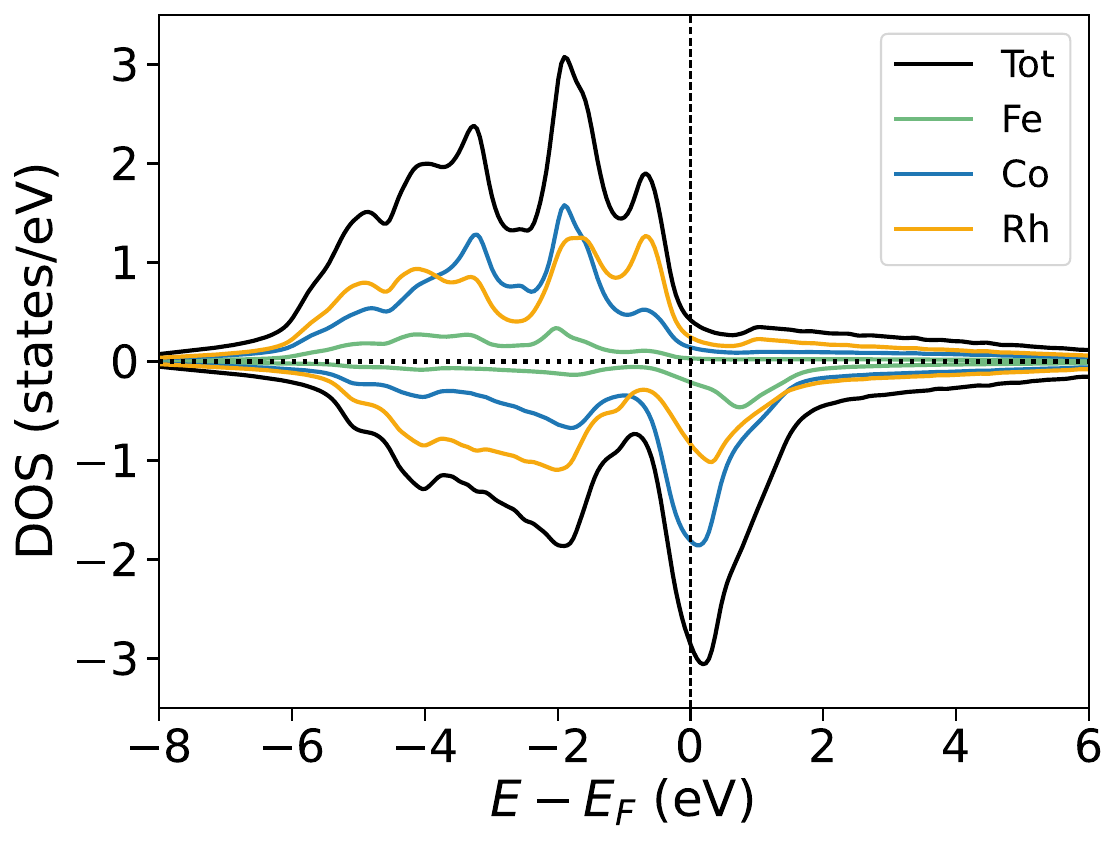}
		\caption{\ce{Fe_{0.2}Co_{0.8}Rh}}
		\label{fig:dos_co80}
	\end{subfigure}
	\caption{Spin-resolved density of states at representative compositions ($x = 0.1, 0.5, 0.8$): Mn-substituted series (top row) and Co-substituted series (bottom row). Mn hole doping progressively shifts $E_F$ out of the majority-spin pseudogap---the majority-spin DOS at $E_F$ rises while the minority-spin DOS falls, with the channels crossing near $x = 0.5$---eroding the spin-channel asymmetry. Co electron doping leaves $E_F$ pinned within the majority-spin pseudogap across the series, preserving the asymmetry and the high spin polarisation. Full $N_\uparrow(E_F)$ and $N_\downarrow(E_F)$ values are listed in \cref{tab:struct-mn,tab:struct-co}.}
	\label{fig:dos}
\end{figure*}

In the Mn series, hole doping induces a rigid-band-like downshift of $E_F$---from 9.82~eV at $x = 0$ to 9.44~eV at $x = 0.5$
(\cref{tab:struct-mn})---that moves it out of the majority-spin pseudogap (\cref{fig:dos_mn10}-\subref{fig:dos_mn80}). The spin-resolved DOS at $E_F$ then responds in opposite directions in the two channels: the majority DOS rises sharply [$N_\uparrow(E_F)$: $0.30 \to 0.75 \to 1.23$~states/eV at $x = 0, 0.5, 0.8$] as $E_F$ descends toward the filled majority $d$-band, while the minority DOS falls [$N_\downarrow(E_F)$: $2.33 \to 0.81 \to 0.50$~states/eV]. The two channels cross near $x \approx 0.5$, where $N_\uparrow \approx N_\downarrow$, so the spin polarisation passes through a compensation point ($P \approx -0.04$ at $x = 0.5$) and reverses sign, from $P \approx -0.77$ at $x = 0$ to $P \approx +0.42$ at $x = 0.8$ (\cref{tab:mnd}). This collapse and reversal of the spin-channel asymmetry reflects a weakening of the effective exchange splitting that stabilizes ferromagnetic order. From a Hume-Rothery band-filling perspective~\cite{hume1939structure,Mizutani2010,Mahat_2022}, the parent compound is stabilized by $E_F$ sitting in the pseudogap (a DOS minimum); Mn hole doping moves $E_F$ out of this minimum, raising the band energy and weakening ferromagnetic stability. This accompanies the competition between ferromagnetic Fe--Fe interactions and the antiferromagnetic interactions introduced by Mn; within the magnetic force theorem this ground-state competition reduces the net exchange coupling and directly underlies the $\sim$450~K suppression of $T_C$ in the Mn series.

% \begin{figure*}[htbp]
% \centering
% \foreach \x in {1,...,8}{%
% \pgfmathtruncatemacro{\mn}{10*\x}
% \pgfmathtruncatemacro{\fe}{100-\mn}
% \pgfmathsetmacro{\stoich}{2-0.02*\x}
% \begin{subfigure}{0.23\textwidth}
% \includegraphics[width=\textwidth]{dos_Mn_\mn.pdf}
% \caption{\ce{Fe_{\fpeval{1-.1*\x}} Mn_{\fpeval{.1*\x}} Rh}}
% \end{subfigure}\hfill%
% }
% \caption{Density of states (DOS) of \mfr~($x = 0.1$--$0.8$). Mn hole doping progressively shifts the Fermi level toward lower energies, out of the majority-spin pseudogap; the majority-spin DOS at $E_F$ rises while the minority-spin DOS falls. This $d$-band filling effect erodes the spin-channel asymmetry and drives the electronic structure toward exchange competition, reflected in the weakening of ferromagnetic exchange at all coordination distances.}
% \label{fig:DOS_Mn}
% \end{figure*}

In stark contrast, Co electron doping shifts $E_F$ only weakly (9.82~eV at $x = 0$ to 9.68~eV at $x = 0.8$;
\cref{tab:struct-co}), keeping it pinned within the majority-spin pseudogap (\cref{fig:dos_co10}-\subref{fig:dos_co80}). The majority $d$-band remains filled, so the majority DOS at $E_F$ stays low [$N_\uparrow(E_F)$: $0.30 \to 0.42$~states/eV] while the added electrons enter the minority $d$-states [$N_\downarrow(E_F)$: $2.33 \to 2.86$~states/eV]. The strong spin-channel asymmetry---and hence the high (minority-character) polarisation $|P| \approx 0.75$ and the large exchange splitting ($\Delta \gtrsim 1$~eV)---is therefore preserved across the entire series (\cref{tab:cod}). This maintains a robust ferromagnetic exchange backbone even as the total moment falls, which is why $T_C$ stays high (836--894~K) across the full range: in this itinerant regime the ordering temperature tracks the exchange splitting and pseudogap pinning, not the size of the total moment. The persistence of $E_F$ within the pseudogap, a Hume-Rothery-type stabilization, suppresses AFM competition and confirms $d$-band filling as the controlling parameter.
\rev{The weak decrease of the tabulated $E_F$ in the Co series does not conflict with the electron-doping picture: the relevant quantity is the position of $E_F$ relative to the DOS features, shown directly on the $E-E_F$ axis of \cref{fig:dos}, where Co doping fills the minority $d$-states while $E_F$ remains pinned in the majority-spin pseudogap.}
%R3.2

% \begin{figure*}[htbp]
% \centering
% \foreach \x in {1,...,8}{%
% \pgfmathtruncatemacro{\co}{10*\x}
% \pgfmathtruncatemacro{\fe}{100-\co}
% \pgfmathsetmacro{\stoich}{2-0.02*\x}
% \begin{subfigure}{0.23\textwidth}
% \includegraphics[width=\textwidth]{dos_Co_\co.pdf}
% \caption{\ce{Fe_{\fpeval{1-.1*\x}} Co_{\fpeval{.1*\x}} Rh}}
% \end{subfigure}\hfill%
% }
% \caption{Density of states (DOS) of \cfr~($x = 0.1$--$0.8$). Co electron doping leaves the Fermi level pinned within the majority-spin pseudogap while filling the minority $d$-states, preserving the spin-channel asymmetry and the high spin polarisation. This electron-filling effect preserves and stabilizes the ferromagnetic electronic structure.}
% \label{fig:DOS_Co}
% \end{figure*}

\subsection{Magneto-volume versus chemical effects}
\label{sec:vol-chem}
\begin{figure*}[htbp]
	\centering
	\begin{subfigure}[b]{0.24\textwidth}
		\centering
		\includegraphics[width=\textwidth]{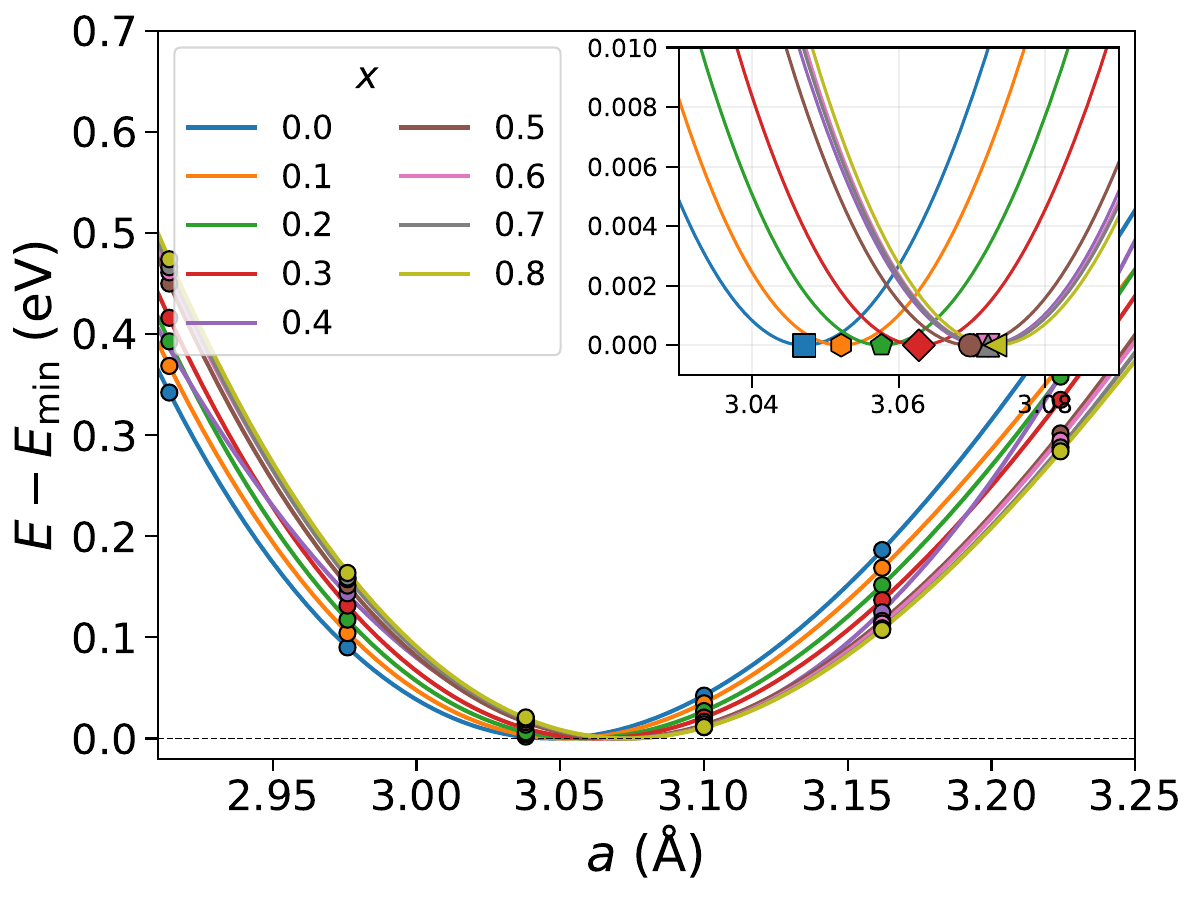}
		\caption{FM \mfr}
		\label{fig:volopt-mnfm}
	\end{subfigure}\hfill
	\begin{subfigure}[b]{0.24\textwidth}
		\centering
		\includegraphics[width=\textwidth]{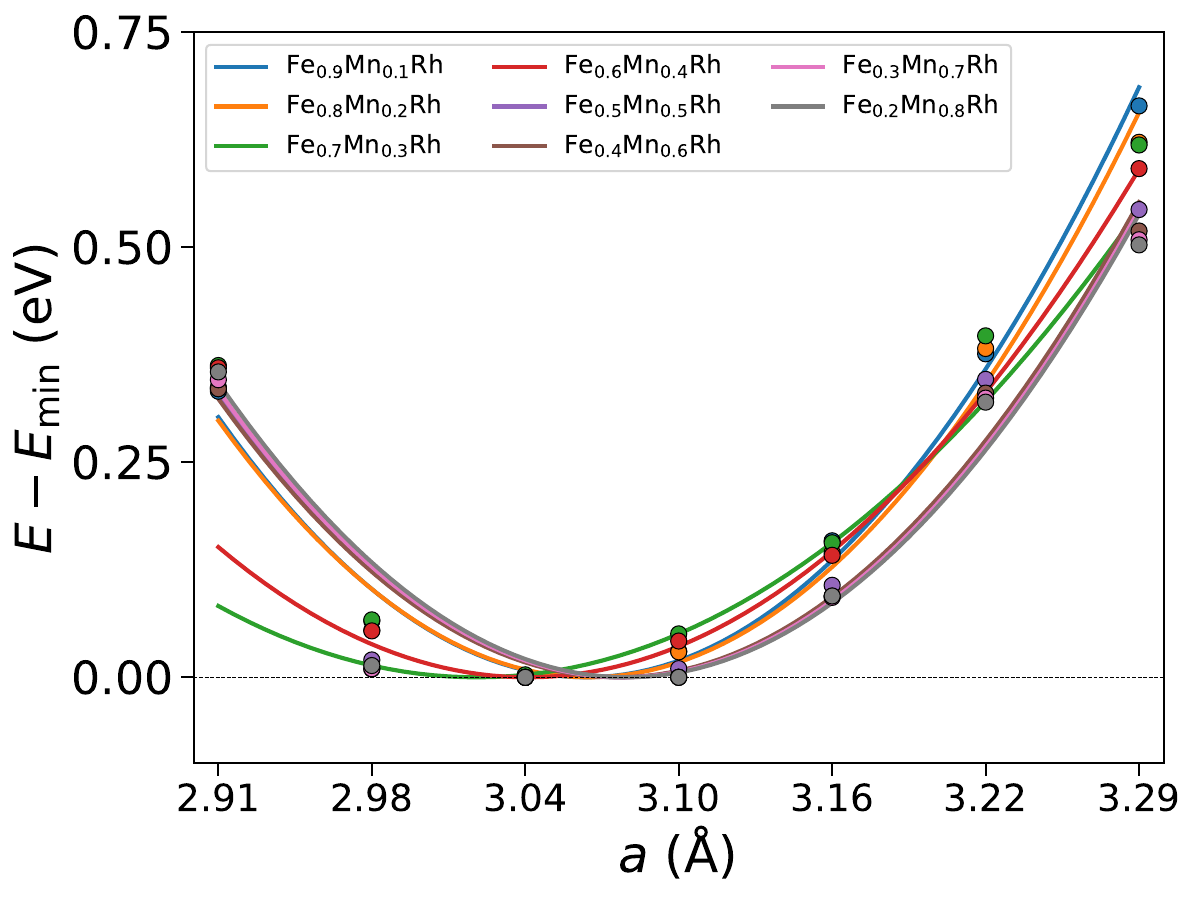}
		\caption{AFM \mfr}
		\label{fig:volopt-mnafm}
	\end{subfigure}
	\begin{subfigure}[b]{0.24\textwidth}
		\centering
		\includegraphics[width=\textwidth]{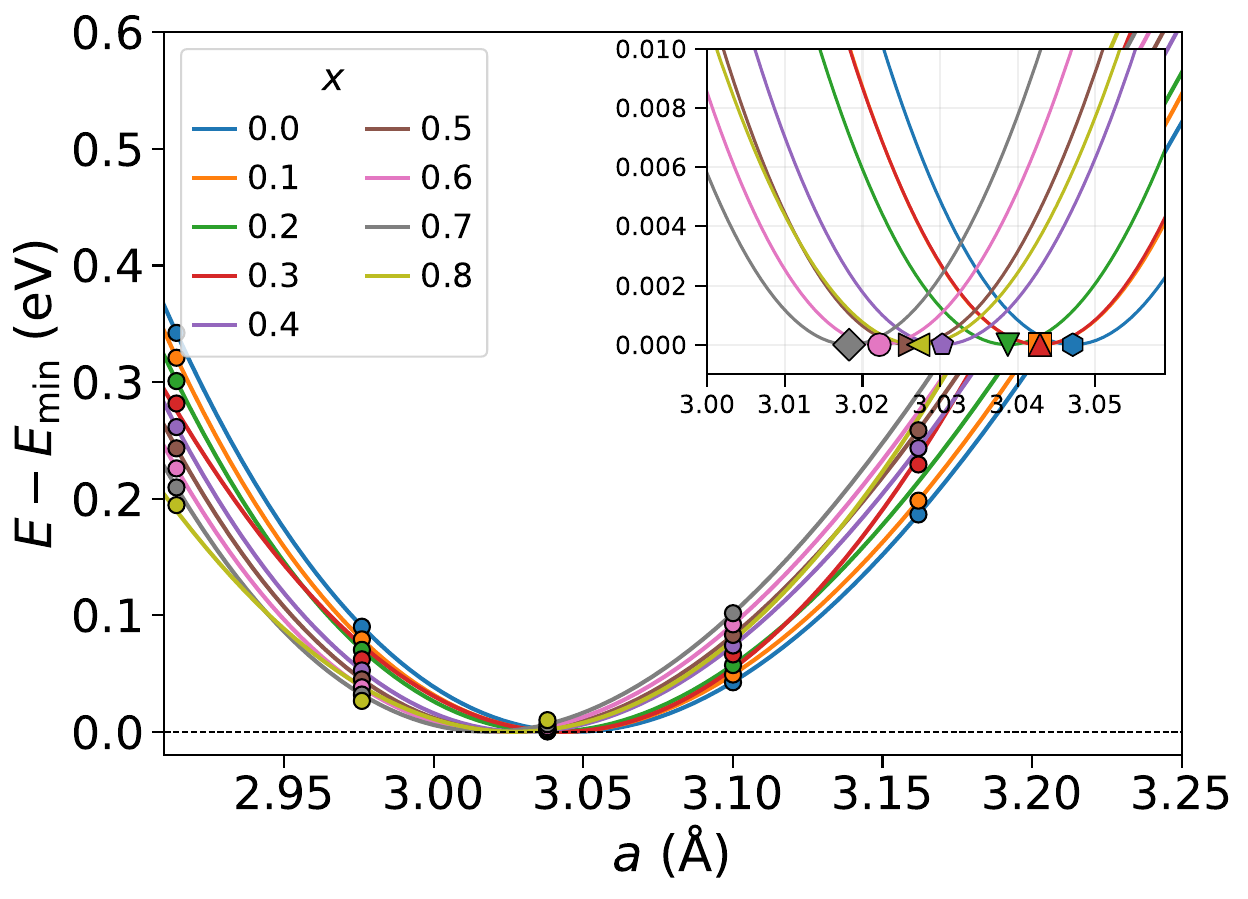}
		\caption{FM \cfr}
		\label{fig:volopt-cofm}
	\end{subfigure}\hfill
	\begin{subfigure}[b]{0.24\textwidth}
		\centering
		\includegraphics[width=\textwidth]{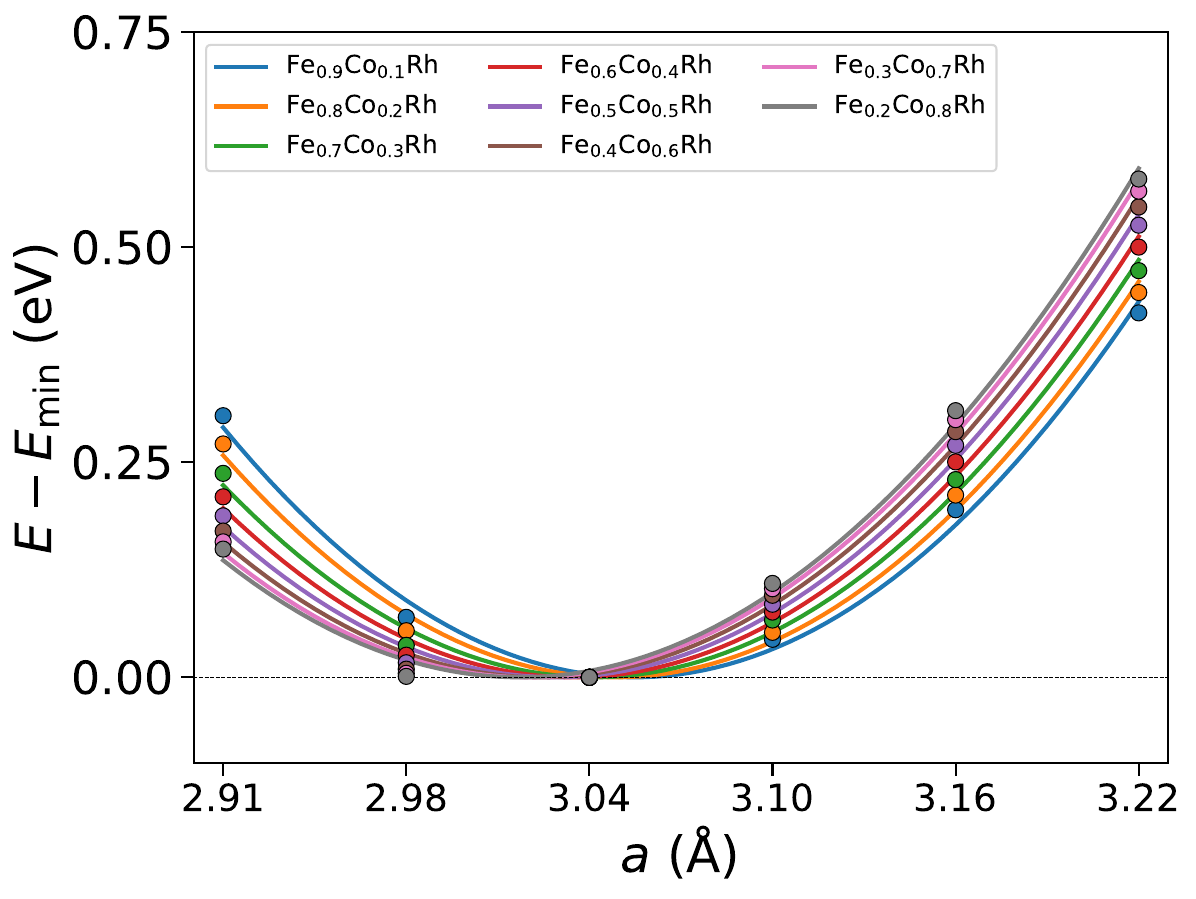}
		\caption{AFM \cfr}
		\label{fig:volopt-coafm}
	\end{subfigure}
	\caption{Total energy versus lattice parameter for the FM and AFM configurations of \mfr~and \cfr~($x = 0.1$--$0.8$). The equilibrium lattice constant for each composition is determined from a quadratic fit to the energy minimum.}
	\label{fig:volopt}
\end{figure*}

The optimised lattice parameters for all compositions are listed in \cref{tab:struct-mn,tab:struct-co} and the corresponding total-energy curves are shown in \cref{fig:volopt}. Across the doped range $x = 0.1$--$0.8$ the Mn lattice parameter changes by only $\sim$0.1\% (3.071 to 3.074~\AA), yet $T_C$ drops by $\sim$450~K. Over the same range the Co lattice parameter contracts by $\sim$0.9\% (3.042 to 3.014~\AA), an order of magnitude more, yet the FM state remains stable with $T_C > 800$~K throughout. If magneto-volume effects were the dominant driver, the Co series---which undergoes the larger volume change---should show the more dramatic magnetic response. Instead the opposite holds: the Mn series, with the smaller volume change but a reduced valence electron count, is the one that destabilises.

% Integrated from SI: full FM/AFM volume-optimization curves (former Fig. S1)

To further isolate the chemical effect, the pure-FeRh reference ($x = 0$, $a = 3.0493$~\AA) provides the appropriate benchmark: its electronic structure and exchange interactions, characterized in our earlier work~\cite{Greeshma_2022}, are robustly ferromagnetic with no sign of competing AFM pathways. This confirms that $d$-band filling---governed by the valence electron count of the substituent ($Z_\text{Mn} = 25$ vs.\ $Z_\text{Co} = 27$ relative to $Z_\text{Fe} = 26$)---rather than the magneto-volume effect, is the primary factor controlling magnetic stability in these alloys.

\subsection{Magnetic moments and spin polarisation}
The redistribution of electronic states above manifests directly in the site-resolved magnetic moments and the net spin polarisation. The values in \cref{tab:mnd,tab:cod} are sublattice-averaged moments; the apparent drop or rise in these tabulated values is a direct consequence of the changing concentration $x$ rather than a collapse of the local atomic magnetism. The \emph{local atomic} moments of the $3d$ constituents are remarkably robust: the host Fe moment ($\approx 3.2$--$3.3~\mu_B$), the Mn site-projected moment ($M_\text{Mn}/x \approx 3.7$--$3.9~\mu_B$), and the Co site-projected moment ($M_\text{Co}/x \approx 2.1~\mu_B$) remain largely stable across the range, indicating a strong localised character governed by on-site Hund's coupling. In contrast, the induced Rh moment depends significantly on composition as the polarizing exchange field from the Fe sublattice is diluted. In the parent \ce{FeRh}, the high total moment ($M_\text{tot} = 4.20~\mu_B$) is supported by a large Fe $d$-state exchange splitting, which induces a substantial Rh moment ($M_\text{Rh} = 0.95~\mu_B$). Within the FM reference used for the Liechtenstein analysis all moments (Fe, Mn/Co, Rh) are parallel, so the signs in \cref{tab:mnd,tab:cod} are positive by construction.

\begin{table*}[ht]
	\begin{center}
		\begin{tabular}{ccccccccc}
			\toprule
			\multirow{2}{*}{$x$} & \multicolumn{4}{c}{Magnetic moment ($\mu_B$)} & \multirow{2}{*}{$P$} & \multirow{2}{*}{$T_C(K)$} &
			\multicolumn{2}{c}{$\eta$}                                                                                                                                            \\\cline{2-5} \cline{8-9}
			                     & Mn                                            & Fe                   & Rh                        & Total  &         &       & Fe     & Mn      \\\midrule

			0.0$^\ast$           & -                                             & 3.2544               & 0.9470                    &
			4.2014               & -0.7710                                       & 833.6                & 0.3222                    & -                                           \\\midrule
			0.1                  & 0.3877                                        & 2.9628               & 0.9387                    & 4.2892 & -0.7273 & 797.1 & 0.1073 & -0.0836 \\
			0.2                  & 0.7706                                        & 2.6346               & 0.9228                    & 4.3279 & -0.6605 & 748.6 & 0.0787 & -0.1287 \\
			0.3                  & 1.1584                                        & 2.3236               & 0.9138                    & 4.3958 & -0.5506 & 692.5 & 0.0547 & -0.1307 \\
			0.4                  & 1.5408                                        & 2.0015               & 0.8899                    & 4.4321 & -0.3405 & 654.2 & 0.0879 & -0.1233 \\
			0.5                  & 1.9143                                        & 1.6718               & 0.8481                    & 4.4342 & -0.0407 & 599.4 & 0.0993 & -0.0947 \\
			0.6                  & 2.2745                                        & 1.3376               & 0.7903                    & 4.4025 & 0.1681  & 531.2 & 0.1102 & -0.0561 \\
			0.7                  & 2.6293                                        & 1.0028               & 0.7201                    & 4.3523 & 0.3129  & 456.7 & 0.1165 & -0.0224 \\
			0.8                  & 2.9859                                        & 0.6683               & 0.6465                    & 4.3007 & 0.4230  & 381.8 & 0.1012 & -0.0230 \\
			\bottomrule
		\end{tabular}
		\caption{Composition-dependent magnetic properties of Mn-substituted FeRh (\mfr), all within the FM reference (parallel alignment). Columns list the spin moments ($\mu_B$) on the Mn, Fe, and Rh sublattices and the total moment per formula unit, followed by the spin polarisation $P$ (\cref{eq:spinpol}), the MFA Curie temperature $T_C$ (\cref{eq:tc}), and the Fe- and Mn-centred exchange competition parameters $\eta$ (\cref{eq:frustration}). The ``Mn'' entry is the \emph{concentration-weighted} sublattice contribution ($x\,M_\text{Mn}$, the per-formula-unit value), so its apparent rise with $x$ reflects increasing Mn content, not a growing local moment: the site-projected local moment $M_\text{Mn}/x$ in fact decreases only slightly, from $\approx 3.88$ to $3.73~\mu_B$. Mn hole doping reduces $T_C$ from 834~K to 382~K and $\eta_\text{Fe}$ from 0.32 to 0.10 across $x = 0$--$0.8$, while $\eta_\text{Mn}$ remains negative throughout. $^{\ast}$The $x = 0$ entry (pure \ce{FeRh}) is discussed in \cite{Greeshma_2022}.}
		\label{tab:mnd}
	\end{center}
\end{table*}

For the Mn series (\cref{tab:mnd}), hole doping drives a non-monotonic total moment: $M_\text{tot}$ rises from $4.20~\mu_B$ ($x = 0$) to a maximum of $4.43~\mu_B$ at $x = 0.50$, then decays to $4.30~\mu_B$ at $x = 0.80$. Despite this modest variation ($\pm 2.6\%$ from the mean), the character of the magnetism changes dramatically, as shown by the spin-polarisation sign change from $P = -0.77$ to $P = +0.42$ across the series (\cref{fig:P}). The progressive loss of Rh-mediated coupling ($M_\text{Rh}$: $0.95 \to 0.65~\mu_B$) together with Fe dilution dominates at high Mn content. This follows directly from the Fermi level moving out of the majority-spin pseudogap (\cref{sec:dos}): the rising majority occupation and falling minority occupation at $E_F$ converge near $x \approx 0.5$, producing a compensation point in the itinerant channel. This collapse of $|P|$ is intrinsically linked to the $T_C$ suppression, as the loss of spin polarisation reflects a reduction in the exchange splitting that stabilizes long-range ferromagnetic order.

\begin{table*}[htpb]
	\begin{center}
		\begin{tabular}{ccccccccc}
			\toprule
			\multirow{2}{*}{$x$} & \multicolumn{4}{c}{Magnetic moment ($\mu_B$)} & \multirow{2}{*}{$P$} & \multirow{2}{*}{$T_C(K)$} & \multicolumn{2}{c}{$\eta$}                                     \\\cline{2-5}\cline{8-9}
			                     & Co                                            & Fe                   & Rh                        & Total                      &         &       & Fe     & Co     \\\midrule

			0.1                  & 0.2089                                        & 2.9226               & 0.9443                    & 4.0758                     & -0.7587 & 836.0 & 0.2160 & 0.3705 \\
			0.2                  & 0.4175                                        & 2.5895               & 0.9344                    & 3.9415                     & -0.7559 & 847.2 & 0.2285 & 0.4335 \\
			0.3                  & 0.6276                                        & 2.2589               & 0.9357                    & 3.8222                     & -0.7509 & 850.7 & 0.2065 & 0.4442 \\
			0.4                  & 0.8396                                        & 1.9329               & 0.9400                    & 3.7126                     & -0.7489 & 855.4 & 0.1795 & 0.4335 \\
			0.5                  & 1.0512                                        & 1.6069               & 0.9439                    & 3.6020                     & -0.7468 & 858.8 & 0.1486 & 0.4096 \\
			0.6                  & 1.2611                                        & 1.2814               & 0.9454                    & 3.4879                     & -0.7459 & 864.3 & 0.1243 & 0.3883 \\
			0.7                  & 1.4692                                        & 0.9574               & 0.9457                    & 3.3723                     & -0.7450 & 874.9 & 0.1080 & 0.3757 \\
			0.8                  & 1.6788                                        & 0.6362               & 0.9505                    & 3.2655                     & -0.7442 & 893.6 & 0.0931 & 0.3702 \\
			\bottomrule
		\end{tabular}
		\caption{Composition-dependent magnetic properties of Co-substituted FeRh (\cfr), all within the FM reference (parallel alignment); columns as in \cref{tab:mnd}. The ``Co'' entry is likewise the concentration-weighted sublattice contribution ($x\,M_\text{Co}$); the site-projected local moment is $M_\text{Co}/x \approx 2.1~\mu_B$, roughly constant. Co electron doping keeps $T_C$ high and weakly increasing (836--894~K) and $\eta_\text{Co} \gtrsim 0.37$ across $x = 0.1$--$0.8$. The spin polarisation remains negative ($P \approx -0.75$), indicating minority-spin dominance at $E_F$ as in the parent compound; ferromagnetism is sustained because the exchange splitting---not the spin-channel balance at $E_F$---sets the coupling strength.}
		\label{tab:cod}
	\end{center}
\end{table*}

In the \cfr~system (\cref{tab:cod}), $M_\text{tot}$ decreases monotonically from $4.08~\mu_B$ ($x = 0.10$) to $3.27~\mu_B$ ($x = 0.80$) as Fe is diluted, yet electron doping holds the Rh moment near $0.94~\mu_B$ across all compositions (variation $<2\%$). The Co atoms contribute a stable local moment ($\approx 2.1~\mu_B$) which, though smaller than the Fe moment it replaces, reinforces the ferromagnetic topology. The spin polarisation stays negative ($P \approx -0.75$, range $-0.76$ to $-0.74$; \cref{fig:P}), indicating minority-spin dominance at $E_F$ as in the parent \ce{FeRh}. Despite the $\sim$20\% reduction in total moment, $T_C$ rises from 836~K to 894~K---direct evidence that the Curie temperature is set by the magnitude of the exchange splitting (and the resulting pseudogap pinning) rather than by the total moment, a hallmark of itinerant magnetism. This preserved exchange, despite the minority-spin Fermi surface, is why Co substitution keeps $T_C > 800$~K across the full range.

\subsection{Magnetic Exchange and Curie Temperature}
\label{sec:jij}
The thermodynamic stability of the magnetic phases in \zfr~is ultimately set by the real-space topology of the interatomic exchange interactions $J_{ij}$. Mapping these as a function of coordination shell traces how $d$-band filling dictates the strength and range of the coupling between the Fe, Rh, and dopant sublattices. In the host FeRh, ferromagnetism is stabilized by a robust ``ferromagnetic backbone'' dominated by nearest-neighbour Fe--Fe and Fe--Rh interactions~\cite{Greeshma_2022}, giving $T_C \approx 834$~K.

\begin{figure*}[htbp]
	\centering
	% Row 1: Mn-centred
	\begin{subfigure}[b]{0.24\textwidth}
		\includegraphics[width=\textwidth]{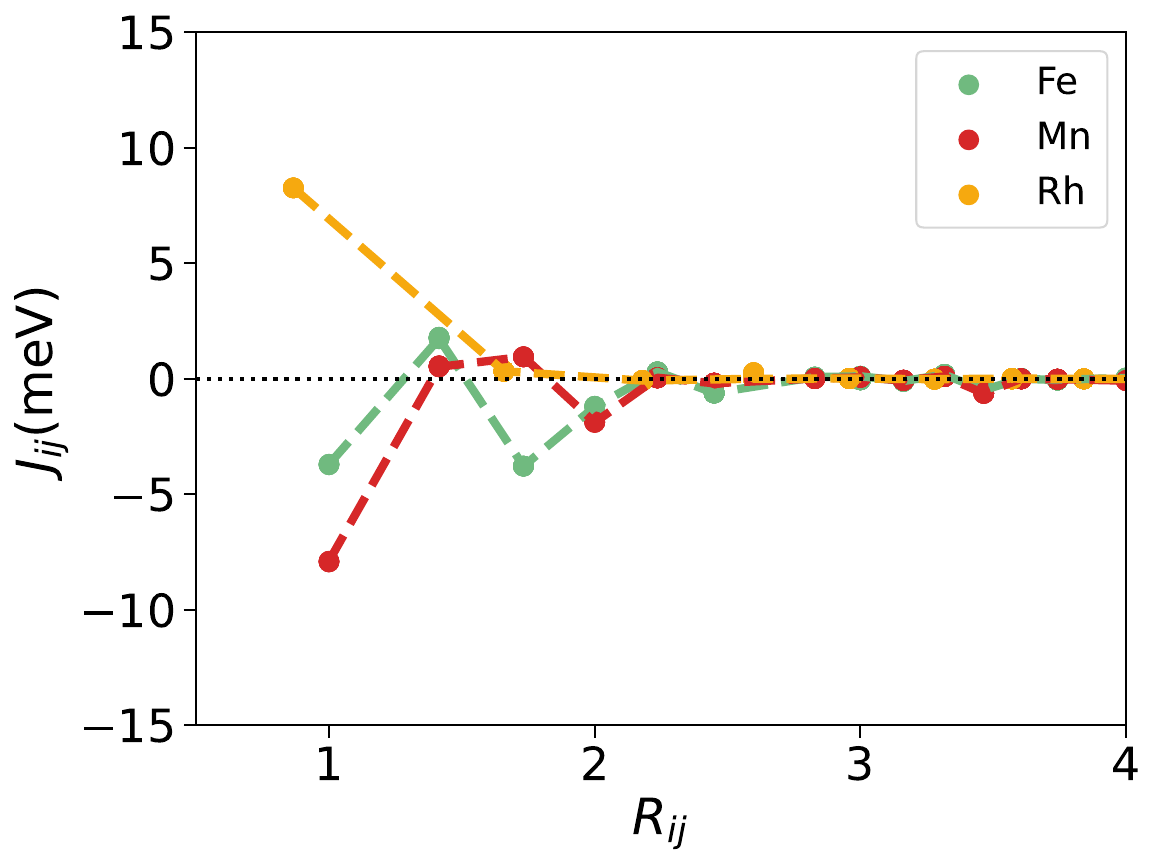}
		\caption{Mn-centred, \ce{Fe_{0.9}Mn_{0.1}Rh}}
		\label{fig:jij_mn10}
	\end{subfigure}\hfill
	\begin{subfigure}[b]{0.24\textwidth}
		\includegraphics[width=\textwidth]{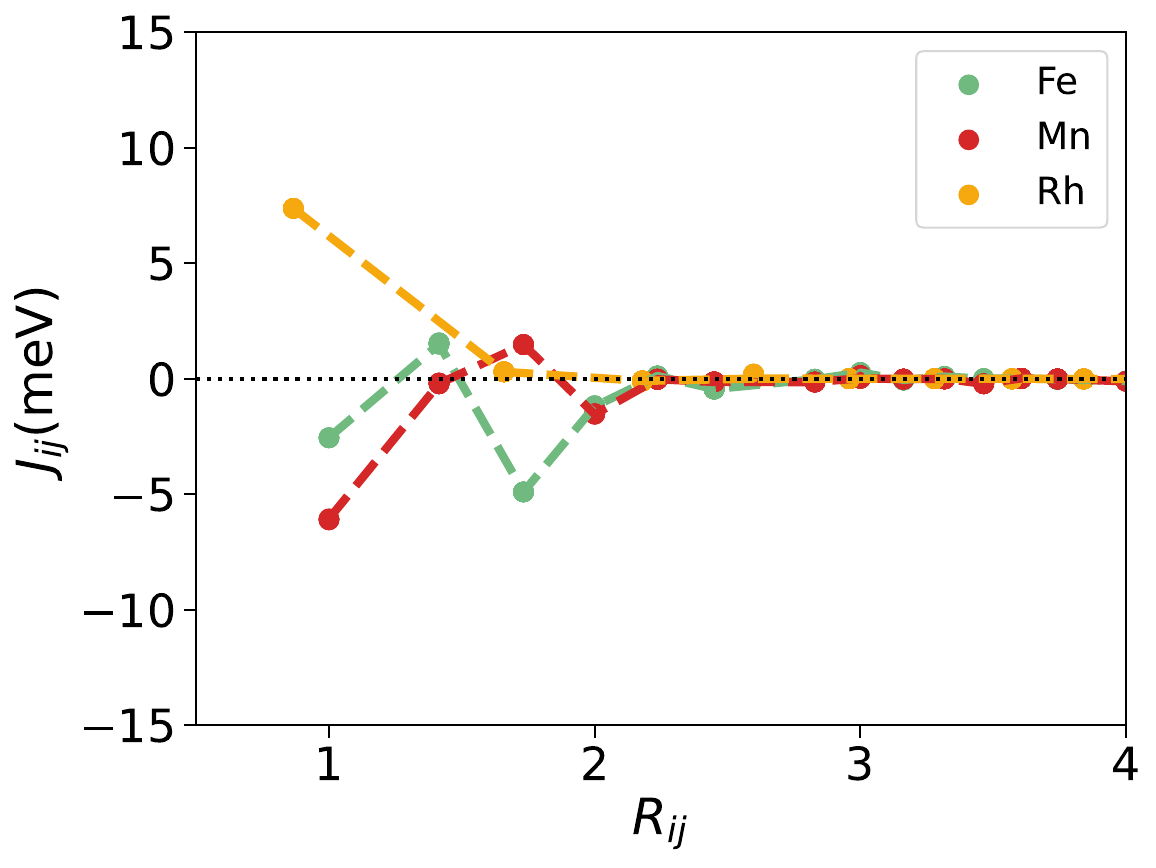}
		\caption{Mn-centred, \ce{Fe_{0.5}Mn_{0.5}Rh}}
		\label{fig:jij_mn50}
	\end{subfigure}\hfill
	\begin{subfigure}[b]{0.24\textwidth}
		\includegraphics[width=\textwidth]{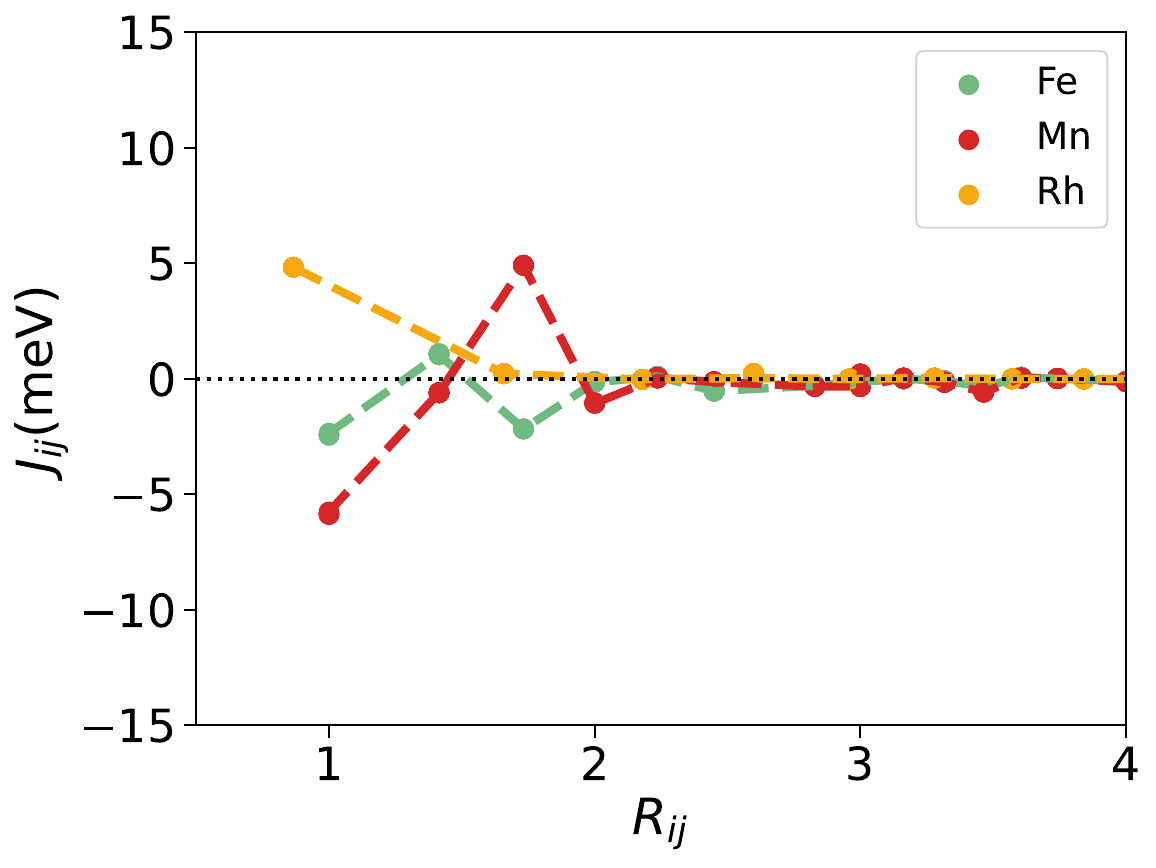}
		\caption{Mn-centred, \ce{Fe_{0.2}Mn_{0.8}Rh}}
		\label{fig:jij_mn80}
	\end{subfigure}
	\begin{subfigure}[b]{0.24\textwidth}
		\includegraphics[width=\textwidth]{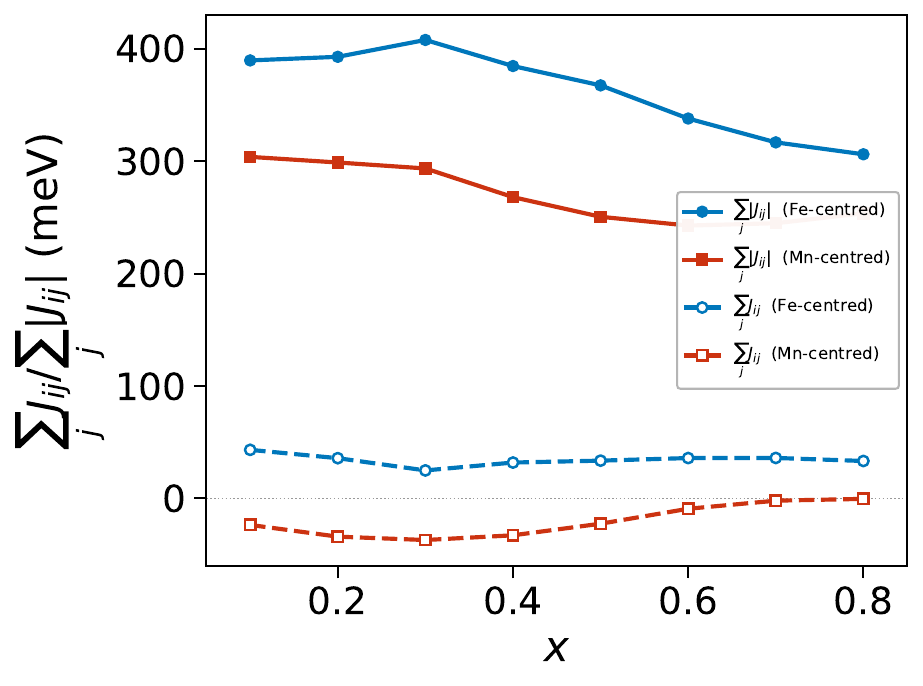}
		\caption{Mn Series}
		\label{fig:sumj_mn}
	\end{subfigure}
	% Row 2: Co-centred
	\begin{subfigure}[b]{0.24\textwidth}
		\includegraphics[width=\textwidth]{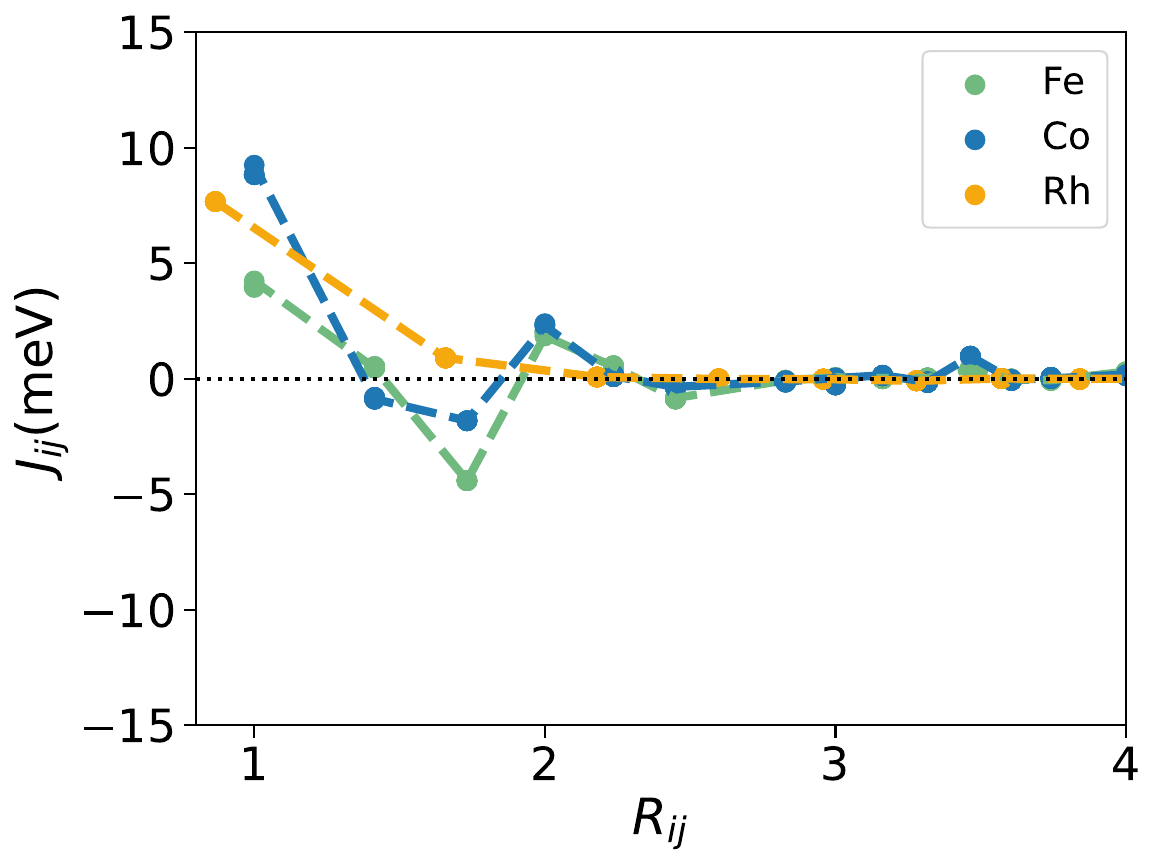}
		\caption{Co-centred, \ce{Fe_{0.9}Co_{0.1}Rh}}
		\label{fig:jij_co10}
	\end{subfigure}\hfill
	\begin{subfigure}[b]{0.24\textwidth}
		\includegraphics[width=\textwidth]{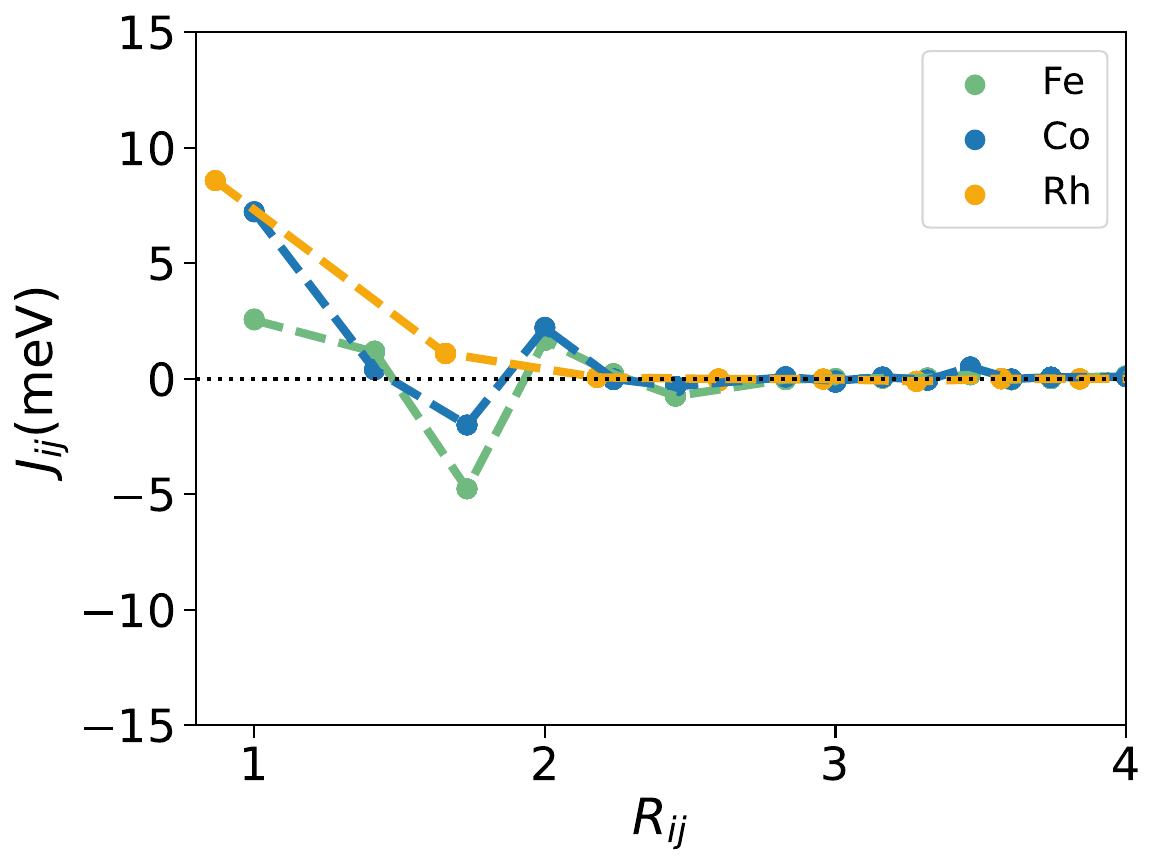}
		\caption{Co-centred, \ce{Fe_{0.5}Co_{0.5}Rh}}
		\label{fig:jij_co50}
	\end{subfigure}\hfill
	\begin{subfigure}[b]{0.24\textwidth}
		\includegraphics[width=\textwidth]{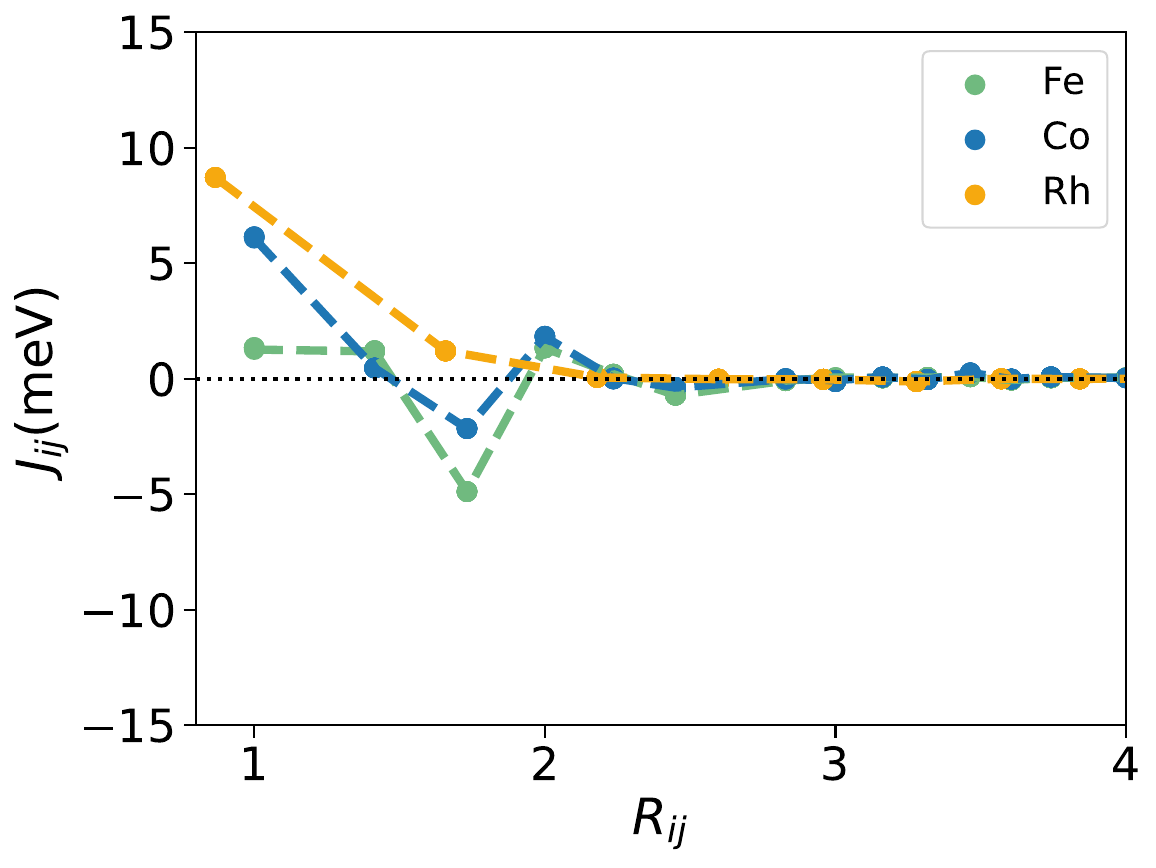}
		\caption{Co-centred, \ce{Fe_{0.2}Co_{0.8}Rh}}
		\label{fig:jij_co80}
	\end{subfigure}
	\begin{subfigure}[b]{0.24\textwidth}
		\includegraphics[width=\textwidth]{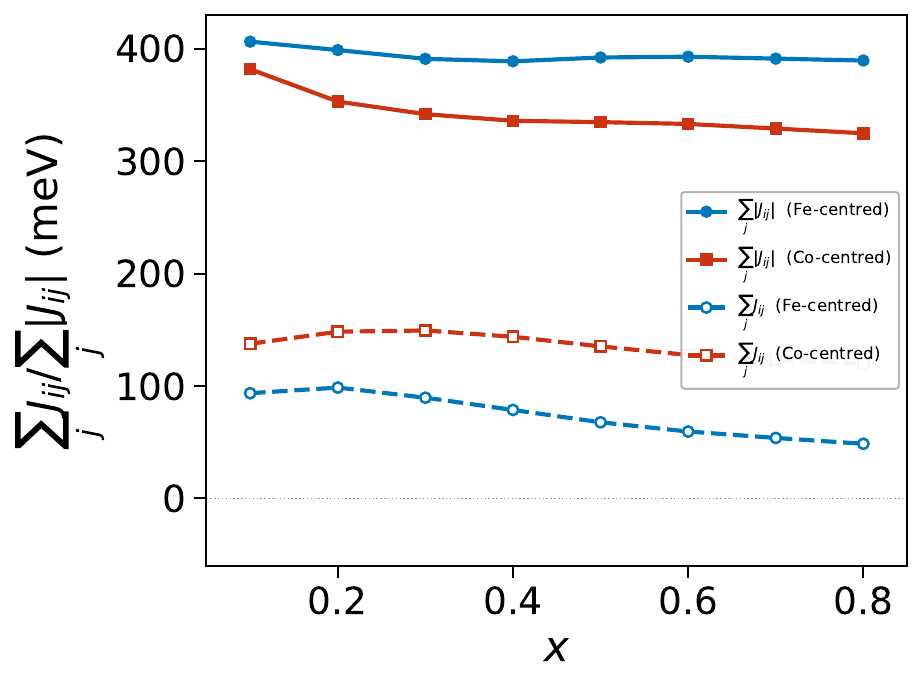}
		\caption{Co Series}
		\label{fig:sumj_co}
	\end{subfigure}
	\caption{Site-resolved exchange interactions $J_{ij}$ versus coordination distance at representative compositions ($x = 0.1, 0.5, 0.8$): Mn-centred (top row) and Co-centred (bottom row). With increasing Mn content the nearest-neighbour ferromagnetic coupling weakens and antiferromagnetic components grow at intermediate shells, producing the near-cancellation ($\eta_\mathrm{Mn} < 0$) that suppresses $T_C$. The Co-centred coupling remains ferromagnetic across all shells and compositions ($\eta_\mathrm{Co} > 0$), preserving the ferromagnetic backbone. The full composition dependence of the net exchange is captured by the $\eta$ panel of \cref{fig:master}.
		\rev{Decomposition of the exchange competition parameter $\eta = \sum_j J_{ij}/\sum_j |J_{ij}|$ into its net ($\sum_j J_{ij}$, dashed, open markers) and total-magnitude ($\sum_j |J_{ij}|$, solid, filled markers) components for the (\subref{fig:sumj_mn})~Mn- and (\subref{fig:sumj_co})~Co-substituted series, each resolved into Fe-centred (blue) and dopant-centred (red) projections. In the Mn series the Fe-centred total magnitude decreases by $\sim$21\% (genuine weakening of the ferromagnetic backbone) while the Mn-centred net coupling collapses toward zero with its magnitude largely preserved (near-perfect FM/AFM cancellation); the two effects cooperate to suppress $T_C$. In the Co series the Fe-centred total magnitude is essentially constant ($\sim$4\% change), confirming that the high $\eta_\text{Co}$ reflects a preserved ferromagnetic backbone. In every case the net coupling is a small fraction of the total magnitude, consistent with $T_C$ being governed by $\sum_j J_{ij}$. Sums are evaluated over the same neighbour shells used for $\eta$ in \cref{tab:mnd,tab:cod}.}}
	\label{fig:jij_rep}
\end{figure*}

Mn substitution systematically destabilizes this network. As hole doping shifts $E_F$ out of the majority-spin pseudogap, the itinerant-mediated exchange pathways are altered: as shown in \cref{tab:mnd} and the Mn-centred panel of \cref{fig:jij_rep}, the Mn-centred interactions ($J_\text{Mn-Fe}$, $J_\text{Mn-Mn}$) develop significant antiferromagnetic components at intermediate shells. This competition is quantified by $\eta$: for the Mn series the Fe-centred $\eta_\text{Fe}$ falls from 0.32 to 0.10 at $x = 0.80$, while $\eta_\text{Mn}$ stays negative ($-0.08$ at $x = 0.10$ to $-0.02$ at $x = 0.80$, \cref{tab:mnd}), so Mn-centred pathways consistently favour antiferromagnetic alignment.
\rev{Because $\eta$ is a normalized ratio, it cannot by itself separate a near-cancellation of strong competing couplings from a
	uniform weakening of all couplings; distinguishing the two requires the net exchange $\sum_j J_{ij}$ and the total exchange
	magnitude $\sum_j |J_{ij}|$ to be examined separately (\cref{fig:sumj_mn}). This decomposition shows that both mechanisms operate in the Mn series. On the Fe-centred pathway $\sum_j |J_{ij}|$ falls by $\sim$21\% (390 to 306~meV between $x = 0.1$ and $0.8$) while the net $\sum_j J_{ij}$ stays small ($\sim$25--43~meV) and nearly constant---a genuine softening of the ferromagnetic backbone. On the Mn-centred pathway the signature is sharper: the net $\sum_j J_{ij}$ collapses almost completely (from $-24$ to $-0.4$~meV) while its magnitude $\sum_j |J_{ij}|$ decreases only modestly ($\sim$16\%, 304 to 254~meV), the hallmark of near-perfect cancellation between ferromagnetic and antiferromagnetic contributions. The $\sim$450~K suppression of $T_C$---to 382~K at $x = 0.80$---therefore reflects two cooperating effects: a real reduction of the total Fe-centred exchange strength and a near-complete cancellation of the net Mn-centred coupling, the latter being the sharper signature and the one that drives $\eta_\text{Mn} \to 0$.}
% This strong-competition regime---where FM and AFM interactions partially cancel---is the microscopic origin of the $\sim$450~K suppression of $T_C$, which falls to 382~K at $x = 0.80$.

% \begin{figure*}[htbp]
% \centering
% \begin{subfigure}[b]{0.48\textwidth}
% \centering
% \includegraphics[width=\textwidth]{jij_mnc_Mn_50.pdf}
% \caption{Mn-centred, \ce{Fe_{0.5}Mn_{0.5}Rh}}
% \label{fig:jij_rep_mn}
% \end{subfigure}\hfill
% \begin{subfigure}[b]{0.48\textwidth}
% \centering
% \includegraphics[width=\textwidth]{jij_coc_Co_50.pdf}
% \caption{Co-centred, \ce{Fe_{0.5}Co_{0.5}Rh}}
% \label{fig:jij_rep_co}
% \end{subfigure}
% \caption{Representative site-resolved exchange interactions $J_{ij}$ versus coordination distance at $x = 0.5$, contrasting the two dopant environments. (\subref{fig:jij_rep_mn})~In the Mn-centred case, the nearest-neighbour ferromagnetic coupling is reduced and antiferromagnetic components appear at intermediate shells, producing the near-cancellation ($\eta_\mathrm{Mn} < 0$) that suppresses $T_C$. (\subref{fig:jij_rep_co})~In the Co-centred case, the coupling remains ferromagnetic across all shells ($\eta_\mathrm{Co} > 0$), preserving the ferromagnetic backbone. The full composition dependence of the net exchange is captured by the $\eta$ panel of \cref{fig:master}.}
% \label{fig:jij_rep}
% \end{figure*}

In contrast, Co substitution reinforces ferromagnetic order through electron doping. By holding $E_F$ within the majority-spin pseudogap, Co preserves the high exchange splitting and ensures that the itinerant electrons mediate ferromagnetic coupling across all relevant shells. As shown in \cref{tab:cod} and the Co-centred panel of \cref{fig:jij_rep}, the Co-alloy exchange topology stays robustly ferromagnetic, with $\eta_\text{Co}$ remaining high across the series---from 0.37 at $x = 0.8$ to a peak of 0.44 at $x = 0.3$.
\rev{The same decomposition confirms that this reflects genuine stabilization rather than a coincidence of the ratio: the
	Fe-centred total exchange magnitude $\sum_j |J_{ij}|$ is essentially constant across the Co series (406 to 390~meV, a $\sim$4\%
	change; \cref{fig:sumj_co}), so the high, nearly flat $\eta_\text{Co}$ tracks a ferromagnetic backbone whose strength is preserved as Fe is replaced. This flat Co baseline is what makes the $\sim$21\% weakening of the Fe-centred magnitude in the Mn series a meaningful, chemistry-specific effect rather than a generic consequence of dilution.}
The absence of significant AFM competition prevents any softening of the magnetic energy landscape and keeps $T_C > 800$~K even at high substitution.

The MFA Curie temperatures, obtained from the $J_{ij}$ spectra, capture the disparity between the two series: the transition from a stable high-$T_C$ ferromagnet (Co) to a thermally fragile magnet with competing exchange (Mn) is set by the presence or absence of dopant-mediated antiferromagnetic pathways in the exchange topology, not by the local moment magnitudes.

\begin{figure*}[htbp]
	\centering
	\foreach \x in {Tc, eta,P}{%
			\begin{subfigure}{0.3\textwidth}
				\includegraphics[width=\textwidth]{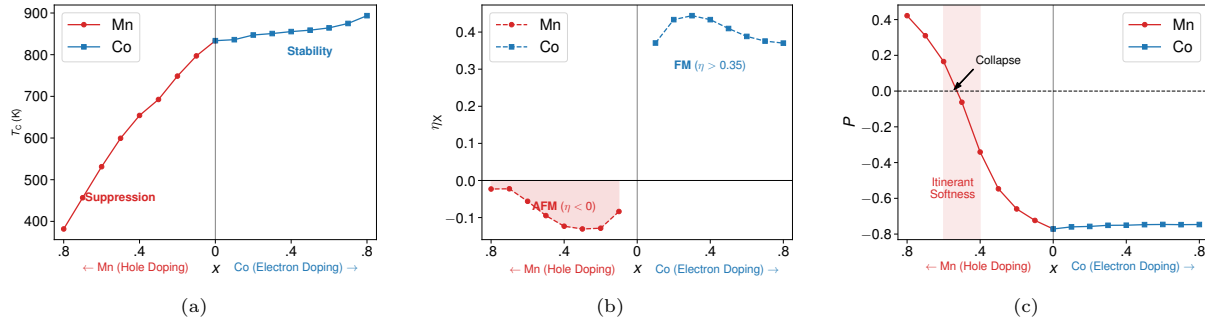}
				\caption{}
				\label{fig:\x}
			\end{subfigure}\hfill%
		}
	\caption{Unified magnetic phase diagram of \zfr~with substitutional disorder on the Fe sublattice. The panels relate the macroscopic stability to its microscopic electronic drivers across the transition from hole-doped (Mn, left) to electron-doped (Co, right) regimes. (\subref{fig:Tc})~Curie temperature $T_C$: Co maintains high $T_C$, whereas Mn drives a $\sim$450~K suppression. (\subref{fig:eta})~Exchange competition parameter $\eta$: the ratio of net to total exchange stays robustly positive for Co ($\eta_\mathrm{Co} \gtrsim 0.37$) but drops below zero for Mn-rich alloys, signalling the onset of antiferromagnetic competition. (\subref{fig:P})~Spin polarisation $P$: the electronic origin of the instability is the collapse of $P$, which crosses zero at $x \approx 0.5$ for the Mn series, in contrast to the pinned, high-polarisation state ($P \approx -0.75$) preserved for Co.}
	\label{fig:master}
\end{figure*}

\section{Discussion}
\label{sec:discussion}

\subsection{$d$-band filling and the electronic mechanism}
The diverging magnetic fates of Mn- and Co-doped FeRh are immediately apparent in the unified phase diagram (\cref{fig:master}): Co substitution maintains a Curie temperature above 800~K, while Mn substitution suppresses $T_C$ to 382~K (\cref{fig:Tc}). The MFA value for pure \ce{FeRh} (834~K) overestimates the experimental Curie temperature of the FM phase ($\approx$675~K)~\cite{Pandey_2025} through the neglect of short-range correlations---and we note this is the FM-phase $T_C$, distinct from the lower metamagnetic AFM--FM transition temperature---but the qualitative bifurcation between the two dopants is robust. This divergence is consistent with the established dependence of magnetic moments on $e/a$ in transition-metal alloys~\cite{Pauling1938,Slater1936,Galanakis2002}, yet our site-resolved $J_{ij}$ analysis exposes the specific exchange-topological mechanism underlying that Slater-Pauling trend. Where early models emphasized magneto-volume instabilities~\cite{Kudrnovsky2015}, the data of \cref{sec:vol-chem} show the $d$-band filling effect to dominate: the Co series undergoes the larger lattice contraction yet stays a robust ferromagnet, whereas the Mn series, with minimal volume change, collapses. This identifies the Fermi-level position relative to the majority-spin pseudogap as the decisive control parameter.

That Mn hole doping shifts $E_F$ out of the majority-spin pseudogap---raising $N_\uparrow(E_F)$ while lowering $N_\downarrow(E_F)$, and thereby eroding the spin-channel asymmetry---accounts for the onset of exchange competition in \cref{fig:eta}, where $\eta$ drops below zero for Mn-centred interactions, signalling that antiferromagnetic pathways have become comparable to the ferromagnetic backbone. A similar destabilization occurs in Mn-based antiperovskites such as \ce{Mn3SnC}, where departures from optimal filling drive a cluster-glass instability~\cite{Gaonkar2020JAC}. The persistently negative $\eta_\mathrm{Mn}$ carries a clear physical meaning: the Mn-centred environment favours antiferromagnetic alignment of the Mn moments relative to the Fe host. Because the magnetic force theorem requires a magnetic reference, we extract $J_{ij}$ about the FM state (the phase realized above the metamagnetic transition); that $\eta_\mathrm{Mn} < 0$ emerges self-consistently from this reference is itself the central result---Mn embedded in the FM host acts as a ``frustrated impurity'' that introduces exchange competition and destabilizes ferromagnetic order, directly producing the $T_C$ suppression.

Conversely, Co acts as a ``magnetic hardener'': electron doping leaves $E_F$ pinned within the majority-spin pseudogap, preserving the large exchange splitting and the strong spin-channel asymmetry. Analogous behaviour is seen in Ni-doped FeRh (isoelectronic to Co), where electron doping stabilizes the ferromagnetic phase against the metamagnetic transition~\cite{Polesya2016}. In our results this appears as $\eta_\mathrm{Co} \gtrsim 0.37$ across the entire Co series (\cref{fig:eta}), consistent with the transport measurements of Seo \textit{et al.}, who found that Co doping enhances the thermal stability of FeRh thin films by suppressing spin-flip scattering~\cite{Seo_2024}.
\rev{
	The microscopic origin of these inter-shell sign changes lies in the itinerant, RKKY-like character of the exchange. Because
	the coupling between the $3d$ moments is mediated by the itinerant electrons---predominantly the broad Rh $4d$ bands---each
	interaction inherits an oscillatory dependence on the inter-site separation, schematically $J(R)\sim\cos(2k_F R+\varphi)/R^3$,
	whose sign at a given coordination shell is fixed by the phase $2k_F R$ set by the spanning vectors of the Fermi surface. This
	sign alternation between shells is directly visible in \cref{fig:jij_rep}, and the cumulative exchange converges only beyond
	several lattice constants, confirming the long-range oscillatory nature of the coupling. As Mn hole doping lowers the $d$-band
	filling and shifts $E_F$ downward relative to the majority-spin pseudogap (\cref{fig:dos,tab:struct-mn}), the Fermi-surface
	geometry---and with it the dominant spanning vector $2k_F$---evolves, shifting the oscillation phase so that antiferromagnetic
	lobes of $J(R)$ move into the intermediate shells that remain ferromagnetic in pure \ce{FeRh}; this is the microscopic origin of
	the inter-shell sign reversal and of the emergence of $\eta_\mathrm{Mn}<0$. For a multiband $d$-electron system this is
	necessarily an effective single-$k_F$ description, but the quantity it rests on---the position of $E_F$ relative to the
	spin-resolved DOS---is precisely the control parameter established throughout this work. The site-resolved sums then separate the
	two physically distinct contributions to this competition (\cref{fig:sumj_co},\subref{fig:sumj_mn}): the Fe-centred projection probes the \emph{global}
	band-filling response of the host backbone, whose total magnitude $\sum_j|J_{ij}|$ weakens by $\sim$21\% across the Mn series,
	while the dopant-centred projection isolates the \emph{local} impurity term, in which the net Mn coupling collapses to near
	zero---a local antiferromagnetic tendency of the kind expected for Mn near half-filling of its $d$ shell. That the Co-centred net
	coupling instead remains large and positive ($\sum_j J_{ij}\approx 120$--$150$~meV, $\eta_\mathrm{Co}\gtrsim 0.37$) confirms that
	this local dopant contribution is element-specific and cleanly separable from the global band-filling trend.
} %R2.2
\subsection{Itinerant magnetic softness and exchange topology}
We use ``itinerant magnetic softness'' to describe the regime identified in the Mn series, defined by three convergent signatures: (i)~the exchange competition parameter $\eta$ approaches or crosses zero, so antiferromagnetic pathways become comparable to the ferromagnetic backbone; (ii)~the spin polarisation $|P|$ collapses, crossing zero at $x \approx 0.5$ (\cref{fig:P}), reflecting a loss of spin-channel asymmetry at $E_F$; and (iii)~the Curie temperature is sharply suppressed. We stress that ``soft'' here is \emph{not} used in the conventional sense of low magnetic anisotropy; it denotes the thermal fragility of long-range ferromagnetic order that arises when competing exchange interactions nearly cancel.

This softness refers to the \emph{exchange topology}, not to the AFM--FM energy difference: over the studied range the FM configuration is energetically well separated from the G-type AFM-II configuration ($\Delta E$ up to $\sim$0.35~eV/atom; \cref{fig:deltaE}). The apparent paradox---a thermally fragile ferromagnet whose FM reference is nonetheless energetically stable against the AFM-II state---is resolved by noting that $T_C$ is set by the net exchange field ($\propto \sum J_{ij}$), a small fraction of the total exchange scale ($\propto \sum |J_{ij}|$).
\rev{The net field is suppressed primarily by near-complete cancellation of the competing ferromagnetic and antiferromagnetic
	couplings---sharpest on the Mn-centred pathway, where $\eta_\text{Mn} \to 0$ as the net $\sum_j J_{ij}$ collapses while the
	individual couplings remain substantial ($\sum_j |J_{ij}|$ falling only $\sim$16\%; \cref{fig:sumj_mn},\subref{fig:sumj_co}). The couplings do not, however, merely cancel: the total Fe-centred exchange magnitude itself weakens by $\sim$21\% across the series, so the ferromagnetic backbone genuinely softens as well. This real reduction of the exchange scale, superimposed on the dominant cancellation, is what makes ``softness'' a literal description of the Mn series rather than merely shorthand for near-cancellation, and both effects together drive the net field below the threshold needed to sustain long-range ferromagnetic order against thermal fluctuations.}% When the ferromagnetic and antiferromagnetic pathways nearly cancel, the net field is too weak to sustain long-range order against thermal fluctuations even though each individual coupling remains substantial. 
This is the ground-state precursor to the enhanced spin fluctuations expected in itinerant systems with competing interactions~\cite{Moriya1985,Mohn1999}. The direction of this trend is consistent with the recent observation by Joshi \textit{et al.} that Mn substitution induces broad thermal hysteresis and a depressed $T_C$~\cite{Joshi2025}; our calculations extend that behaviour across the composition range and trace its electronic origin to the approach toward half-filling.

The mapping underlying this analysis is the Liechtenstein formula~\cite{Liechtenstein_1984}, which is not a localized-spin model: it is a perturbative force-theorem scheme~\cite{Liechtenstein1987,Turek_2006,Sasioglu2025} that retains the full itinerant electronic structure through the scattering-path operators $\hat{T}^{ij}_\sigma$ and is justified wherever well-defined local moments exist, as we verified for all Fe, Mn, Co, and Rh sites. The three convergent signatures together provide a self-consistent fingerprint of itinerant magnetic softness that is qualitatively distinct from localized-moment frustration: in classical frustrated magnets the competing interactions arise from geometric constraints on a fixed spin lattice, whereas here the competition is generated \emph{electronically}, by the redistribution of spectral weight across spin channels as the $d$ band is filled. This distinction is practically useful---because the softness is governed by a continuous electronic parameter, the valence electron count, it can in principle be graded by compositional design, offering a route to itinerant ferromagnets with prescribed thermal sensitivity of the magnetic order.
\rev{Three considerations bound the scope of the FM-reference analysis adopted here. First, semi-local functionals such as PBE render the FeRh AFM--FM energy balance only approximately and are known to favour the ferromagnetic state~\cite{Kudrnovsky2015}; we therefore do not interpret the absolute magnitude of $\Delta E$ as a quantitative ground-state prediction. Our conclusions rest instead on the exchange topology and on the composition trends in $T_C$, $\eta$, and $P$---all evaluated internally to the FM reference and governed by the motion of $E_F$ relative to the majority-spin pseudogap---which are robust to a rigid offset of the FM--AFM energy difference. Second, the Liechtenstein $J_{ij}$ describe transverse spin rotations and do not capture the longitudinal (amplitude) response of the moments across the metamagnetic transition; this caveat is most relevant for the Rh moment, which is induced by the surrounding Fe environment (\cref{tab:mnd,tab:cod}) and may soften in magnitude rather than behave as a rigid local moment. Third, the competing antiferromagnetic pathways that emerge at high Mn content ($\eta_\mathrm{Mn} < 0$) indicate that magnetic configurations beyond the collinear FM state---including, but not restricted to, the G-type AFM-II ordering---may become energetically competitive in the Mn-rich regime; a full mapping of that competition would require explicit total-energy comparison of several non-collinear and antiferromagnetic orderings and lies beyond the present scope. None of these qualifications alters the central finding: that $d$-band filling controls the exchange topology and thermal stability of the FM phase across the studied range.}%R2.1 + R3.1 + R3.3 

\section{Conclusion}
\label{sec:conclusion}

We have established that the magnetic stability of B2-ordered FeRh alloys with substitutional disorder on the Fe sublattice is governed by $d$-band filling relative to the majority-spin pseudogap. Mapping the \zfr~phase space ($x = 0.1$--$0.8$) with first-principles CPA calculations and site-resolved Liechtenstein exchange analysis, we show that the Fermi-level position provides a systematic, microscopic framework for the exchange topology and the resulting magnetic stability.

Mn substitution (hole doping) shifts $E_F$ out of the majority-spin pseudogap, eroding the spin-channel asymmetry: the majority-spin DOS at $E_F$ rises while the minority-spin DOS falls. This produces a ``magnetic softening''---competing antiferromagnetic pathways emerge ($\eta_\mathrm{Mn} < 0$), the spin polarisation collapses and reverses sign (from $|P| \approx 0.77$ to $P \approx +0.42$), and the Curie temperature falls from 834~K to 382~K. This finite-temperature instability sets in even though the FM configuration remains energetically well separated from the G-type AFM-II configuration over the studied range ($\Delta E$ up to $\sim$0.35~eV/atom), confirming that it is driven by \rev{the exchange topology i.e. the near-cancellation of competing $J_{ij}$ compounded by a genuine weakening of the total exchange magnitude,} rather than by AFM--FM energy proximity. In stark contrast, Co substitution (electron doping) acts as a ``magnetic hardener'': by leaving $E_F$ pinned within the majority-spin pseudogap it preserves the large exchange splitting, maintaining ferromagnetic coupling ($\eta_\mathrm{Co} \gtrsim 0.37$) and high spin polarisation ($|P| \approx 0.75$) across the full range. The magneto-volume analysis (\cref{sec:vol-chem}) confirms that these divergent trends follow the valence electron count rather than the lattice parameter.

These results show that the magnetic evolution reflects specific electronic-structure tuning via $d$-band filling rather than mere sublattice dilution. The itinerant magnetic softness of the Mn series is a continuously tunable feature, distinct from localized-moment instability, and suggests a route to functional magnets where thermal sensitivity of the ferromagnetic order is desired, as in magnetocaloric or spintronic applications. \rev{Concretely, a continuously tunable $T_C$ can be positioned near a target operating temperature, where the magnetization and hence the isothermal entropy change exploited in magnetocaloric refrigeration varies most steeply with temperature, and where the same proximity enables temperature-gated switching of the magnetic state in thermally-assisted spintronics.}%R1.4
\revB{Because the mechanism rests only on the position of $E_F$ relative to a spin-split pseudogap, it is not specific to FeRh and should operate in other itinerant magnets whose Fermi level lies near a pseudogap of one spin channel.}%R1.1
Natural extensions are the experimental validation of the predicted $T_C$ suppression in Mn-doped thin films and the application of this CPA framework to other $3d$ substituents (Cr, V) and to disordered A1/A2 structures, to map the limits of $d$-band-filling control in itinerant ferromagnets.

\section*{CRediT authorship contribution statement}
\textbf{Greeshma R:} Investigation, Data curation, Formal analysis, Visualization, Writing -- original draft.
\textbf{Rudra Banerjee:} Conceptualization, Methodology, Supervision, Resources, Writing -- review \& editing.

\section*{Acknowledgements}
We express our sincere gratitude to the high performance computing centre (HPCC), SRMIST and the Department of Physics and
Nanotechnology for their support of the computational facility. We are appreciative of the computational resource provided by the
Param Smriti supercomputing facility under National Supercomputing Mission (NSM, CDAC), Government of India. We also express our
heartiest gratitude to Selective Excellence Research Initiative (SERI), SRMIST for their support in our research work. We have
used LLM for clean-up and correction of text.

\section*{Declaration of competing interest}
The authors declare that they have no known competing financial interests or personal relationships that could have appeared to influence the work reported in this paper.

\section*{Data availability}
Data will be made available on request.
% \bibliographystyle{unsrt}
% \bibliography{biblio}

\end{document}